\documentclass[iop]{emulateapj}

\usepackage{amsmath} 
\usepackage{natbib}
\setcitestyle{round,aysep={},yysep={;}}

\usepackage{graphicx}
\usepackage{color}
\newcommand{\gsim}{\raisebox{-0.13cm}{~\shortstack{$>$ \\[-0.07cm]
      $\sim$}}~}
\newcommand{\lsim}{\raisebox{-0.13cm}{~\shortstack{$<$ \\[-0.07cm]
      $\sim$}}~}

\newcommand{\be}{\begin{equation}}
\newcommand{\ee}{\end{equation}}

\shorttitle{Age of massive galaxies at redshift 8}
\shortauthors{L\'opez-Corredoira et al.}

\begin{document}

\title{Age of massive galaxies at redshift 8}
\author{M. L\'opez-Corredoira$^{1,2,3}$, F. Melia$^{4,5}$, 
J.-J. Wei$^{6,7}$, C.-Y. Gao$^{6,7}$}
\altaffiltext{1}{Instituto de Astrofisica de Canarias, E-38205 La Laguna, 
Tenerife, Spain; martin@lopez-corredoira.com}
\altaffiltext{2}{PIFI-Visiting Scientist 2023 of Chinese Academy of Sciences 
at PMO$^{6}$ and National Astronomical Observatories, Beijing 100012, China}
\altaffiltext{3}{Departamento de Astrofisica, Universidad de La Laguna, E-38206 
La Laguna, Tenerife, Spain}
\altaffiltext{4}{Department of Physics, The Applied Math Program, and Department 
of Astronomy, The University of Arizona, AZ 85721, USA}
\altaffiltext{5}{John Woodruff Simpson Fellow}
\altaffiltext{6}{Purple Mountain Observatory, Chinese Academy of Sciences, 
Nanjing 210023, China}
\altaffiltext{7}{School of Astronomy and Space Sciences, University of Science 
and Technology of China, Hefei 230026, China}

\begin{abstract}
Recent James Webb Space Telescope (JWST) data analyses have shown that massive 
red galaxies existed at redshifts $z\gsim 6$, a discovery that is difficult to 
understand in the context of standard cosmology ($\Lambda $CDM). Here we analyze 
these observations more deeply by fitting a stellar population model to the optical 
and near-infrared photometric data. These fits include a main stellar population in 
addition to a residual younger population and with the same extinction for both (a 
lower extinction for the younger population is unphysical). Extra stellar populations 
or the inclusion of an AGN component do not significantly improve the fits. These 
galaxies are being viewed at very high redshifts, with an average $\langle z\rangle 
\approx 8.2$, when the $\Lambda$CDM Universe was only $\approx 600$ Myr old.  This 
result conflicts with the inferred ages of these galaxies, however, which were on 
average between 0.9 and 2.4 Gyr old within 95\% CL. Given the sequence of star 
formation and galaxy assembly in the standard model, these galaxies should instead 
be even younger than 290 Myr on average, for which our analysis assigns a probability
of only $<3\times 10^{-4}$ ($\gtrsim 3.6\sigma $ tension). This outcome 
may indicate the need to consider non-standard cosmologies. Nevertheless, our
conclusions result from several approximations in stellar astrophysics and extinction, 
so they should be taken with a grain of salt. Further research is necessary to 
corroborate the possible existence of galaxies older than the $\Lambda $CDM universe
at their observed redshifts.
\end{abstract}


\section{Introduction}\label{.intro}
Recent James Webb Space Telescope (JWST) data analyses have shown that massive 
red galaxies existed at redshifts $z\gsim 6$, a discovery that is difficult to 
understand in the context of standard cosmology ($\Lambda $CDM) \citep{Lab23,Boy23}.
Here we investigate the age of these galaxies, providing additional evidence of 
an unavoidable controversy if they are old, as we shall demonstrate in this paper, 
especially if they are older than the presumed age of the Universe at the
time corresponding to their redshifts.

There is a broad literature on the topic of the old age of massive high-z galaxies, 
and the tension this produces for standard theories of galaxy formation 
\citep[Sect. 7.2]{Lop22}. Recent JWST observations have discovered very high redshift
galaxies, up to 17, with derived estimates of some of their ages \citep{Mel23}.
The high density of these galaxies has been called the `impossibly early galaxy' 
problem, based on the detection of several orders of magnitude more massive haloes 
at very much higher redshift than predicted \citep{Ste16}. Furthermore, the large 
rotational velocity and large content of cold gas in some high redshift galaxies 
remain a challenge to efforts of reproducing the standard predictions of galaxy
formation \citep{Nee20}. 

The usual methods for determining the galaxies' age include SED (Spectral 
Energy Distribution) fitting or spectral analyses. Spectral analyses may provide more 
accurate measurements, but require very large exposure times, so they are available only 
for a few galaxies and are importantly affected by the age-metallicity degeneracy 
in those cases of low signal/noise \citep{Car03}. Nonetheless, there have been some
useful studies in this direction. For instance, \citet{Sch06} used the strength of 
H$_\delta $ for $z\sim 0.9$ galaxies and the fit of the whole spectrum in their models. 
Their approach seems also appropriate for our galaxies with redshift up to 1.2. A more 
accurate method of age determination, such as the use of H$_\gamma $ \citep{Vaz99,Yam06}, 
would need a very high signal/noise in the spectra, which is not reachable for high redshift 
galaxies.  However, \citet{Gla17,Sch18} were able to analyze the spectra of quiescent galaxies 
from the FourStar Galaxy Evolution Survey (ZFOURGE) using the combined H$_\beta$+H$_\gamma$+H$_\delta $
absorption lines at $3<z<4$ to infer stellar ages $\lesssim 1$ Gyr. Another possible age 
determination is given through some break. \cite{Spi97} used the breaks at 
2640 \AA \ and 2900 \AA \ and the fit of the whole spectrum to determine that the age 
of a galaxy at $z=1.55$ was larger than 3.5 Gyr.
Balmer break has been used in JWST high-z galaxies with different results \citep{Ste23,Vik24}. 
But the Balmer break alone cannot break the dust-age degeneracy, 
so a more global analysis of the spectral energy distribution is required.

By pushing our observations to redshifts larger than six,
we are contributing here to an exploration of this long-standing and deepening 
crisis that has thus far eluded any clear resolution. In this paper,
we shall use the technique of SED fitting because, for the available spectra
of JWST high $z$ galaxies \citep[e.g.,][]{Sch22,Cur23,Vik24}, the resolution 
and signal/noise of absorption lines are not enough to permit a determination
of their ages. In particular, for the 13 galaxies selected by \citet{Lab23}, there are
only four of them with spectra (see Table \ref{Tab:bestfits}) with significant signal/noise only for
some emission lines, thus they do not permit the analysis of absorption lines or breaks in the 
continuum.


\begin{table*}
\caption{Best fit results with two SSPs. Errors represent the limits 
for which the templates of GALAXEV fit the data within 95\% CL ($\equiv 2\sigma $) 
(within the resolution of the templates). The ages are expressed in Gyr. Redshifts 
of id. 13050, 28984, 35900, 39575 are fixed to their spectroscopic [S] values 
(https://dawn-cph.github.io/dja/spectroscopy/nirspec/). References for the
spectroscopic redshifts: \cite{Koc23}: [Koc23]; \cite{Fuj23}: [Fuj23].}
\begin{center}
\begin{tabular}{cccccccccc}
Galaxy ID & $z$ & $\log _{10}[{\rm age_{\rm old}}]$ & $\log _{10}[{\rm age_{\rm young}}]$ & $A_2$ & $[M/H]$ & $A_V$ & $\chi ^2_{\rm red}$ 
\\ \hline
2859 &  $9.85^{+0.87}_{-0.50}$ & $0.40^{+0.64}_{-0.59}$ & $-2.30^{+1.30}_{-0}$ &  $0.02^{+0.05}_{-0.01}$ &
$+0.4^{+0}_{-0.8}$ & $0^{+1.5}_{-0}$ & 1.69 \\
7274 &  $9.87^{+1.33}_{-3.87}$ &   $0.70^{+0.34}_{-1.24}$ & $-2.30^{+1.30}_{-0}$ &  $0.03^{+0.48}_{-0.02}$ &
$-0.4\pm 0$ & $0.07^{+1.43}_{-0.07}$ & 29.39 \\
11184 &  $7.18^{+0.47}_{-0.39}$ &   $-0.54^{+0.34}_{-0.46}$ & $-1.00^{+0.29}_{-1.30}$ &  $0.48^{+0.16}_{-0.47}$ &
$0^{+0}_{-0.4}$ & $0^{+0.75}_{-0}$ & 11.62 \\
13050 &  $5.62$ [Koc23] & $1.04^{+0.00}_{-2.64}$ & $-2.30^{+3.00}_{-0.00}$ &  $0.51^{+0.21}_{-0.51}$ &
$+0.4^{+0}_{-0.8}$ & $2.05^{+0.95}_{-2.05}$ & 7.47 \\
14924 &  $9.16^{+1.04}_{-0.76}$ &   $1.04^{+0}_{-1.58}$ & $-1.60^{+0.60}_{-0.70}$ &  $0.12^{+0.24}_{-0.06}$ &
$+0.4^{+0}_{-0.8}$ & $0.31^{+0.79}_{-0.31}$ & 4.47 \\
16624 &  $9.83^{+1.97}_{-1.03}$ &   $1.04^{+0}_{-2.04}$ & $-2.30^{+1.30}_{-0}$ &  $0.09^{+0.27}_{-0.07}$ &
$+0.4^{+0}_{-0.8}$ & $0.45^{+0.55}_{-0.45}$ & 8.26 \\
21834 &  $9.87^{+0.56}_{-0.52}$ &   $1.04^{+0}_{-1.24}$ & $-1.60^{+0.60}_{-0.70}$ &  $0.06^{+0.09}_{-0.03}$ &
$+0.4^{+0}_{-0.8}$ & $0.26^{+0.74}_{-0.26}$ & 0.39 \\
25666 &  $6.83^{+3.17}_{-1.83}$ &   $1.04^{+0}_{-2.64}$ & $-1.60^{+2.30}_{-0.70}$ &  $0.22^{+0.38}_{-0.22}$ &
$+0.4^{+0}_{-0.8}$ & $0.10^{+1.90}_{-0.10}$ & 24.00 \\
28984 &  $7.09$ [S] &   $1.04^{+0}_{-1.09}$ & $-2.30^{+1.30}_{-0}$ &  $0.19^{+0.41}_{-0.15}$ &
$+0.4^{+0}_{-0.8}$ & $0.32^{+1.68}_{-0.32}$ & 4.51  \\
35300 &  $7.77$ [Fuj23]  &   $1.04^{+0}_{-1.24}$ & $-2.30^{+1.76}_{-0.00}$ &  $0.27^{+0.33}_{-0.24}$ &
$+0.4^{+0}_{-0.8}$ & $1.25^{+0.85}_{-1.25}$ & 1.81 \\
37888 &  $6.98^{+3.02}_{-1.98}$ &   $1.04^{+0}_{-2.64}$ & $-1.60^{+2.30}_{-0.70}$ &  $0.70^{+0.29}_{-0.67}$ &
$-0.4\pm 0$ & $0.51^{+1.49}_{-0.51}$ & 7.68  \\
38094 &  $8.10^{+0.62}_{-2.10}$ &  $-0.54^{+1.58}_{-0.46}$ & $-2.30^{+2.11}_{-0}$ &  $0.02^{+0.58}_{-0.02}$ &
$0^{+0}_{-0.4}$ & $0.42^{+1.58}_{-0.42}$ & 142.92  \\
39575 &  $7.99$ [Fuj23] &  $1.04^{+0.00}_{-2.64}$ & $-2.30^{+2.26}_{-0}$ &  $0.62^{+0.24}_{-0.59}$ &
$-0.4\pm 0$ & $1.49^{+0.61}_{-1.49}$ & 3.39  \\ \hline
Stacked: all & --- &  $1.04^{+0}_{-0.90}$ & $-2.30^{+0.70}_{-0}$ &  $0.31^{+0.29}_{-0.21}$ &
$+0.4^{+0}_{-0.8}$ & $1.06^{+0.94}_{-1.06}$  & 6.34 \\
Stacked: 4 $z_{\rm spec}$ & --- &  $1.04^{+0}_{-0.90}$ & $-2.30^{+0.70}_{-0}$ &  $0.26^{+0.34}_{-0.17}$ &
$+0.4^{+0}_{-0.8}$ & $0.89^{+1.11}_{-0.89}$ & 4.08  \\ 
St.a,$\lambda _r\le 5000$\AA  & --- &  $1.04^{+0}_{-1.09}$ & $-1.60^{+0.40}_{-0.70}$ &  $0.53^{+0.16}_{-0.25}$ &
$-0.4\pm 0$ & $0.75^{+0.31}_{-0.75}$  & 0.56 \\
St.4,$\lambda _r\le 5000$\AA  & --- &  $1.04^{+0}_{-1.09}$ & $-2.30^{+0.70}_{-0}$ &  $0.24^{+0.36}_{-0.18}$ &
$+0.4^{+0}_{-0.8}$ & $0.72^{+1.28}_{-0.72}$ & 4.51  \\ 
St.a,$\lambda _r\le 4000$\AA  & --- &  $1.04^{+0}_{-1.24}$ & $-1.60^{+0.40}_{-0.70}$ &  $0.18^{+0.54}_{-0.04}$ &
$+0.4^{+0}_{-0.8}$ & $0.38^{+0.72}_{-0.26}$  & 0.37 \\
St.4,$\lambda _r\le 4000$\AA  & --- &  $1.04^{+0}_{-2.64}$ & $-2.30^{+0.70}_{-0}$ &  $0.29^{+0.46}_{-0.19}$ &
$+0.4^{+0}_{-0.8}$ & $0.75\pm 0.75$ & 1.24 \\ \hline
\end{tabular}
\end{center}
\label{Tab:bestfits}
\end{table*}


\section{Data}\label{.data}
With the first observations by the JWST Cosmic Evolution Early Release Science 
(CEERS) program, multiband photometry at 1-5$\mu $m (filters F115W, F150W, F200W, 
F277W, F356W, F410M, F444W), together with preexisting Hubble Space Telescope 
(HST) photometry in the visible (filters F435W, F606W, F814N), a catalog of 42\,729 
sources over an $\approx 38$ arcmin$^2$ field of view has been produced. Within 
this catalog, galaxies have been selected based on the criteria that they were not 
detected in the HST filters [SNR(F435W, F606W, F814N)$<2$], but have a double break 
around the Lyman-$\alpha $ and Balmer spectral regions correponding to redshifts 
$7\lsim z\lsim 9$ [F150W-F277W$<0.7$; F277W-F444W$>1.0$], with good SNR 
[F444W$<27$ AB; F150W$<29$ AB; SNR(F444W)$>8$]. These criteria were 
chosen: (i) to ensure no secondary redshift solutions at low-$z$, and (ii) to 
filter massive galaxies with high M/L ratios. A manual inspection ensured the 
absence of any artefacts in the images. In total, a subsample of 13 galaxies was 
obtained by \citet{Lab23}, the data we shall use throughout this paper.

The conversion of $F_\nu $ into $F_\lambda $ considers the shape of the transmission 
curves, $T(\lambda )$, of the filters:
$F_\lambda (erg\,s^{-1}cm^{-2}\AA^{-1})=2.998\times 10^{-14}F_\nu(nJy)\times 
\frac{\Delta \nu}{\Delta \lambda }$, $\Delta \lambda=\frac{[\int d\lambda \, 
T(\lambda )]^2}{\int d\lambda \,T^2(\lambda )}$ 
$\Delta \nu=\frac{[\int d\nu \,T(\nu )]^2}{\int d\nu \,T^2(\nu )}$.

Four of these galaxies have spectroscopic redshifts. For the remaining nine,
we calculate photometric redshifts that are quite reliable: spectroscopic
and photometric redshifts agree within the error bars and there is a very low
probability of including interlopers in the catalog \citep{Fin23}, as confirmed
for the four galaxies with spectroscopic redshifts in our sample \citep{Koc23,Fuj23}.

\section{SED fitting with two SSPs}

\subsection{Method}
\label{.methods1}

\begin{figure*}
\vspace{0cm}
\centering
\includegraphics[width=8cm]{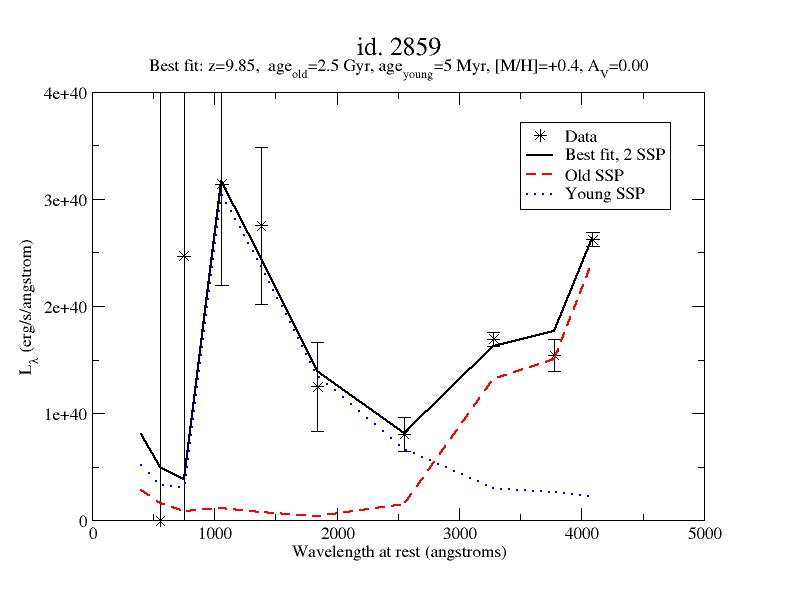}
\hspace{.2cm}
\includegraphics[width=8cm]{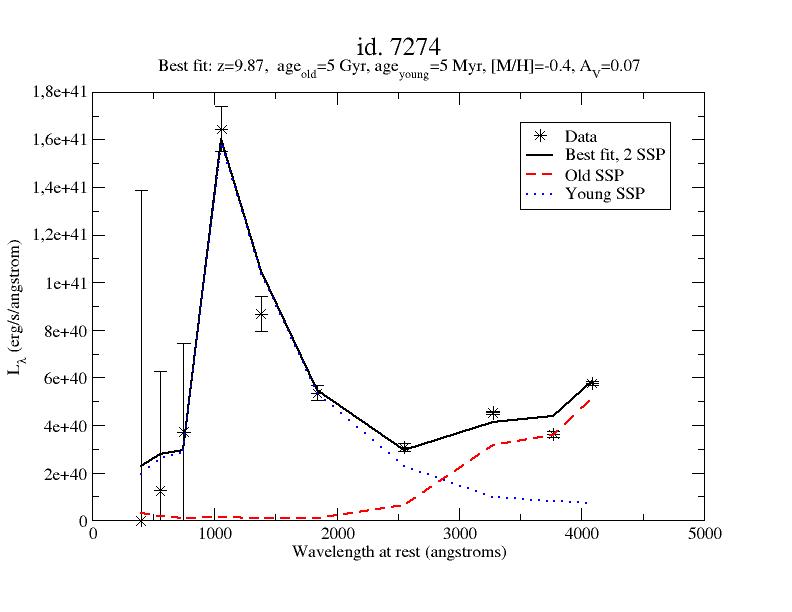}
\vspace{.2cm}
\includegraphics[width=8cm]{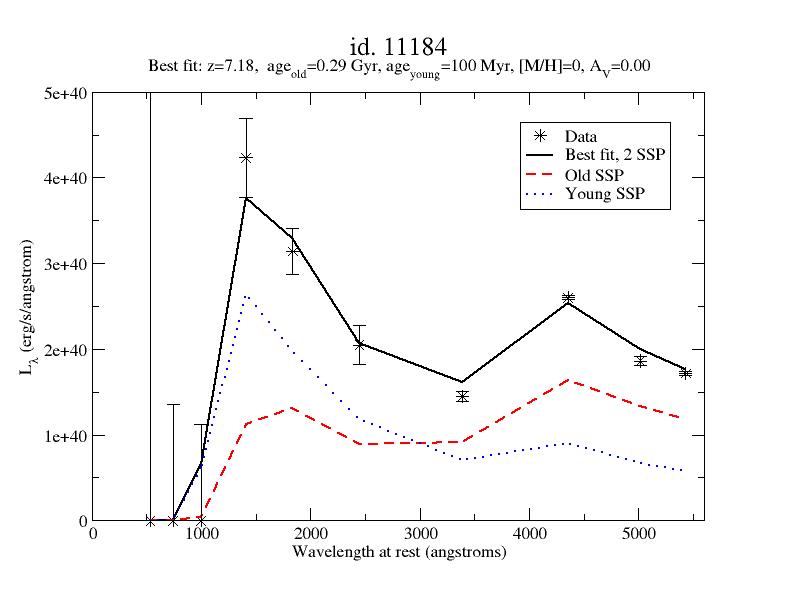}
\hspace{.2cm}
\includegraphics[width=8cm]{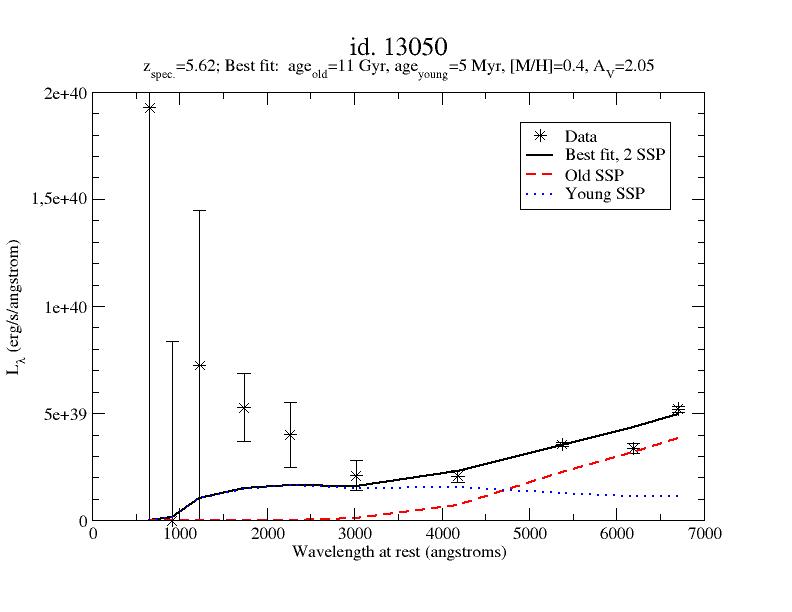}
\vspace{.2cm}
\includegraphics[width=8cm]{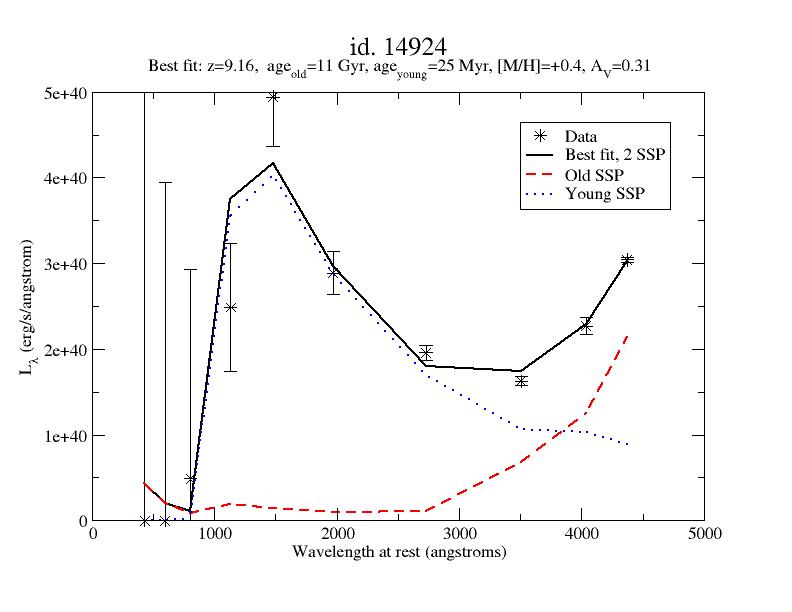}
\hspace{.2cm}
\includegraphics[width=8cm]{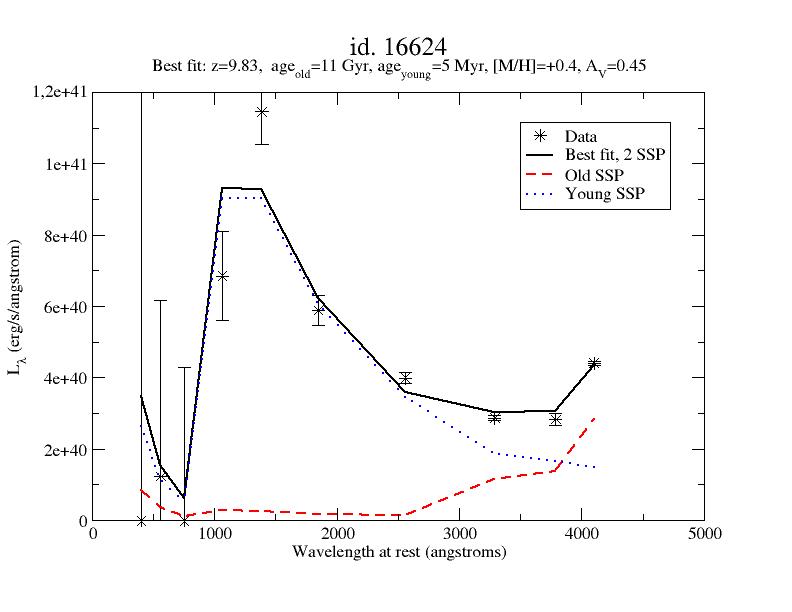}
\vspace{.2cm}
\vspace{.2cm}
\caption{The best fit SEDs of the 13 galaxies in our sample, shown here at 
rest assuming the redshift inferred from the fitting. Negative luminosities 
are represented with a value of zero and its corresponding error bar.}
\label{Fig:sedr}
\end{figure*}

When we plot the SED  of these 13 sources using 
$F_\lambda $ vs. $\lambda $ instead of $F_\nu $ vs. $\lambda $, we clearly see 
the characteristic V-shape of a double-aged population in every case 
(Fig.~\ref{Fig:sedr}). Note that the $x$-axis represents the rest wavelength,
i.e., $\lambda _{\rm observed}/(1+z)$, where $z$ is derived via the fitting described
below, but this conversion of the spectrum into the rest frame does not affect its 
shape. An alternative selection of JWST high-$z$ galaxies within the
catalog of extremely red objects (EROs) is also dominated by V-shaped SEDs \citep{Bar24}: 
JWST-CEERS EROs show blue color in the short-wavelength NIRCam bands (F150W - F277W $\sim $ 0), 
even without requiring a priori any constraint around the Lyman-break.

We cannot explain these SEDs with a single stellar population (SSP), either 
young or old. The fact that there exists a valley between the two high peaks around 
the Lyman-$\alpha $ and Balmer breaks, with a minimum at $\lambda _{\rm rest}\sim 
3000$ \AA\ , indicates that we need at least two SSPs to model the photometry \citep{Lop17,Gao24}: 
a young one (very luminous at far-UV, but with a very low stellar mass contribution: 
$<5$\% in the sample of \citet{Lop17}; and we will see in \S \ref{.results}/Table \ref{Tab:masses} 
that the best fits give ratios $\le 1$\%, though with wide errors, for 12 of 13 galaxies treated 
here) to account for the first peak, and an older one to explain the second. A single SSP of 
$\sim 100$ Myr may produce a double peak too, but 
with very low amplitude, which is not enough to reproduce the conspicuous V-shape 
we observe. Exponentially decaying, extended star formation models improve the 
fits slightly with respect to just one SSP model, but they are considerably worse 
than fits with two SSPs, further supporting the residual star formation scenario 
\citep{Lop17}. In Appendix \ref{.lephare}, we corroborate this result 
with the present sample. This remaining contribution from a young, likely residual, 
star formation component with an age $\le$100 Myr, is consistent with an in situ 
formation scenario, likely produced by a more recent episode of gas accretion.  

The galaxies we model show this V-shaped SED profile, and are massive and red with no 
indication of a large extinction (otherwise we would not see the far-UV component), 
typical of relatively old quiescent component-dominated structures; as we shall see,
they are well described, on average, by two SSPs. In our analysis, one of the SSPs
is relatively old, while the second is young. We fit the free parameters by minimizing 
the reduced chi-square. In total, we use six free parameters, aside from the 
amplitude. These are: the redshift ($z$, assumed to be $\ge 5$; spectroscopic when 
available and fitted otherwise), the ratio of old/young population ($A_2$), 
the age of the old population (age$_{\rm old}$), the age of the young population 
(age$_{\rm young}$), the metallicity [M/H] (the same for both populations; indeed, 
it is not important what the metallicity is for the younger component because the 
models are almost completely insensitive to it), and the internal extinction in the 
V-filter at rest ($A_V$; assumed to be $\le 3.0$). The age-metallicity-dust 
degeneracy \citep{Con13} may appear when we have little information on the photometry, 
for instance some few colors; in the selected filters, however, we are able to break 
this degeneracy. In particular, the Balmer break is affected very little by the 
extinction and is mostly dependent on age, with some small dependence on metallicity. 
In any case, the error bars of our fits indicate the level of this degeneracy. The 
fit of the data using a model with $f$ free parameters (in our case here, $f=7$) is 
carried out with a convential minimization of $\chi^2$ to the theoretical fluxes:
\begin{equation}
\label{Ftheor}
F_{\rm theor.}(\lambda _i)=\frac{L_0}{4\pi d_L(z)^2(1+z)}\times
\end{equation}\[
[\langle L_{\rm SSP}
({\rm age _{\rm old},[M/H]}, A_V;\lambda /(1+z))\rangle _T
\]\[
+A_2\langle L_{\rm SSP}
({\rm age _{\rm young},[M/H]}, A_V;\lambda /(1+z))\rangle _T]
\,,\]
where $\langle (...) \rangle _T=\frac{\int d\lambda \,(...) T(\lambda)}
{d\lambda \,T(\lambda )}$; $d_L(z)$ is the luminosity distance, assuming 
$\Lambda $CDM with $H_0=70$ km s$^{-1}$ Mpc$^{-1}$ and $\Omega _\Lambda=0.7$. 
Note that this conversion of flux into luminosity needs a cosmological model, 
but the SED fitting is totally independent of the cosmology; $d_L$ affects 
only the amplitude of the luminosity, not its shape used in the SED fitting.  

The theoretical spectra $L_0L_{\rm SSP}$ are derived using the GALAXEV 
stellar population synthesis model \citep{Bru03}, which computes the spectral 
evolution of stellar populations from 91 \AA \ to 160 $\mu $m at rest. The 
model incorporates all of the stellar evolutionary phases and faithfully 
reproduces the observed optical and near-infrared color-magnitude diagrams 
of the Galactic star clusters with various ages and metallicities, and 
typical galaxy spectra from the Sloan Digital Sky Survey (SDSS) \citep{Bru03}. 
It was also used to fit high-redshift galaxies with results comparable to 
those of other stellar population synthesis models \citep{Lop17,Gao24}. We adopt 
the 30 templates (ten ages: 0.005, 0.025, 0.10, 0.29, 0.64, 0.90, 1.4, 2.5, 
5.0, 11 Gyr; three metallicities: [M/H]=0, $\pm 0.4$) of instantaneous-burst 
models used to fit the continua in the SDSS galaxy spectra \citep{Tre03}. 
The normalization of the spectra is chosen so that $L_{\rm SSP} (\lambda =
$5500 \AA )=1. This means that $A_2$ represents the ratio of young/old rest 
luminosity at 5,500 \AA .

We note that the stellar population synthesis model of 
\citet{Bru03} contains emission lines in the calculation of the average 
fluxes of each SSP, in cases where we found a very young galaxy with 
significant star formation. In principle, we do not expect young galaxies 
because the criterion we use to select our galaxies, following \citet{Lab23}, 
chooses old and massive galaxies (though with a residual component of a
young population). Other authors, e.g., \citet{End23}, used a very different 
selection of galaxies, with faint UV (no Lyman-$\alpha$ break) and lacking 
a strong Balmer break, and found a significant representation in their
samples of very young galaxies with star formation, but this is not our 
case here, as we shall see.

For the extinction law, we use an attenuation curve often utilized in high-redshift 
studies: Calzetti's law \citep{Cal00}, derived empirically for a sample of nearby 
starburst (SB) galaxies containing small magellanic cloud-like dust grains. 
The application of this law is suggested for the central star-forming regions of galaxies, 
and is therefore appropriate for high-$z$ galaxies, though some authors allow for
the possibility that the JWST high-z galaxies may have different dust properties 
\citep{Mar23}. For this reason, Calzetti's law is frequently used to correct the 
inferred SFR at high-redshift, and we use this extinction law here with the parameter 
$R_V=4.05$. This law is valid for $\lambda >1200$ \AA \ at rest; for lower wavelengths, 
we use the theoretical model with $R_V=4.0$ by \citet{Wei01} normalized at 
$\lambda =1200$ \AA \ from Calzetti's law.

The models EAZY used by \citet{Lab23} that provide the best fits 
(according to their Fig. 3) have been produced using a template set by removing 
the oldest stars, which is similar to other models that use universe priors. We 
should not use any priors. We want to test whether the ages are compatible with 
standard cosmology, so we cannot assume a priori that they are younger than the 
Universe. And using only one single stellar population cannot simultaneously
explain both the Lyman-$\alpha$ and Balmer breaks.  Moreover, in order to
produce their fits within a limited maximum age, \citet{Lab23} combined a dusty 
stellar population (producing the reddening around the Balmer break) with a young 
stellar population (producing the Far-UV Lyman break) without, or very little, dust.\footnote{This
is shown explicitly in the first version of the arXiv preprint (https://arxiv.org/abs/2207.12446v1), where
$F_\lambda $ is plotted instead of $F_\nu$ (and the V-shape is more evident with $F_\lambda $),
and the top-right panel of Fig. 3 clearly shows how they use a dusty stellar population
to explain the Balmer break, and this dust does not take part in the far-UV range.} 
This practice of fitting V-shaped SEDs at high-$z$ as a combination of a 
young, low-mass, low-attenuation galaxy (i.e., a typical Lyman-break galaxy) and, on 
the other, a more massive and dusty galaxy, is quite common \citep[e.g.,][]{Bar24}.
But this is not physical, because the extinction of dust (if any) should be applied to 
the whole stellar population, especially the youngest component that produces the 
Far-UV break, given that the dust is more abundant in young populations \citep{Mal22}.
At a minimum, extinction in the young population cannot be lower than the extinction 
in the old population. Here we apply the same extinction to both the old and young 
populations.

Once we identify the free parameters that minimize the previous 
expression, their corresponding error bars are derived by constraining the models 
that follow \citep{Avn76}
\begin{equation}
\label{Avni}
\chi^2_{\rm red}<\chi^2_{\rm red,\, minimum}
\left[1+\frac{f(CL,N_{\rm dof})}{N_{\rm dof}}\right]
.\end{equation}
For seven free parameters and $N=10$ ($N_{\rm dof}=$3 degrees of freedom), 
$f(CL)=7.81$ at $2\sigma $; for the case of six free parameters and $N=10$ 
(when the redshift is fixed; 4 degrees of freedom), $f(CL)=9.49$ at $2\sigma $; 
for 6 free parameters  and $N=13$ (in the stacked SED; 7 degrees of freedom), $f(CL)=8.18$ at 
$1\sigma $, $f(CL)=14.1$ at $2\sigma $. The inclusion of the factor 
$\chi^2_{\rm red,\,minimum}$ on the right-hand side of the inequality 
takes into account the fact that the errors were underestimated or overestimated 
when $\chi^2_{\rm red,\, minimum}\ne 1$; that is, this is equivalent to assuming 
that our fit is a ``good fit'' (with $\chi^2_{\rm red,\, minimum}\approx 1$) and 
multiplying the error bars by some factor to get it. For the average 
stacked SED, we also calculate confidence levels through a maximum likelihood 
algorithm or direct probability of $\chi ^2$ (see Appendix \ref{.maxlh}).

There are many publicly available algorithms of SED fitting \citep{Pac23}. 
Here we use our own algorithm instead of one of these in order to maintain better control
over the different parameters being fitted: using two SSPs, using the whole range of ages 
instead of limiting them to be younger than the Universe, establishing global extinction 
rather than adding combinations of extinctions and ages that are unphysical, and
deriving the error bars for the fitted parameters. In Appendix \ref{.lephare}, we carry 
out some calculations using one of the publicly available algorithms---{\it le PHARE} 
package---to show consistency between our results and those based on other independent 
calculations.

\begin{figure*}
\includegraphics[width=8cm]{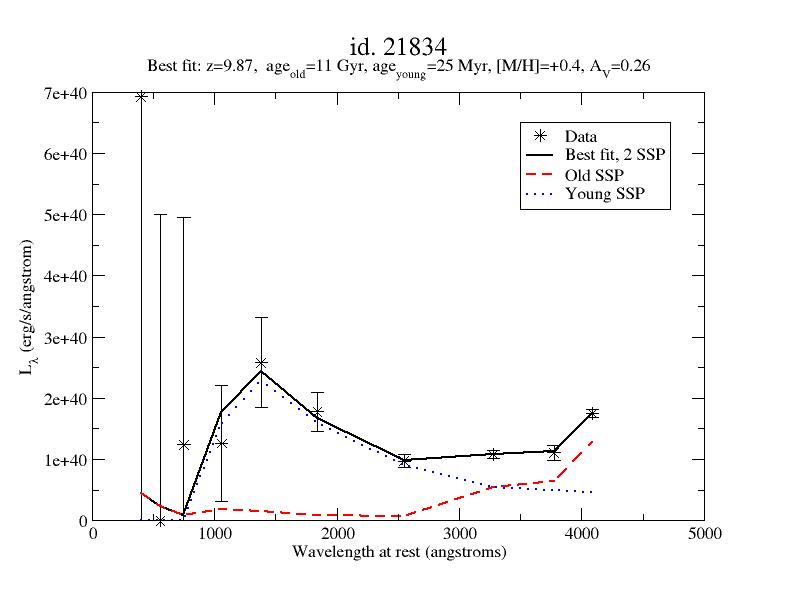}
\hspace{.2cm}
\includegraphics[width=8cm]{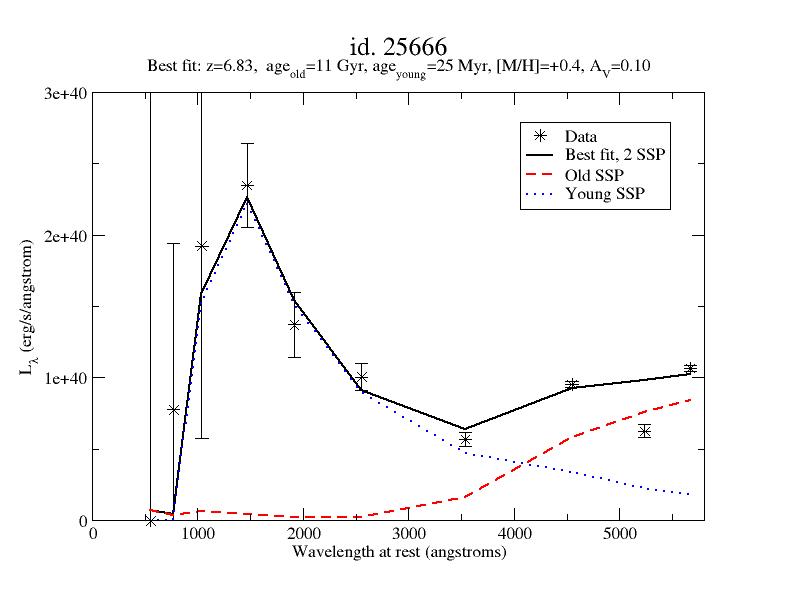}
\vspace{.2cm}
\includegraphics[width=8cm]{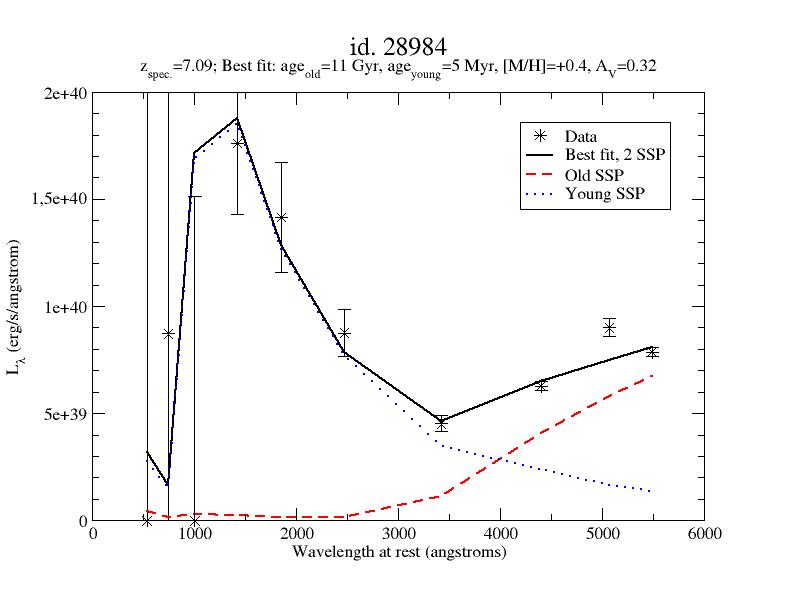}
\hspace{.2cm}
\includegraphics[width=8cm]{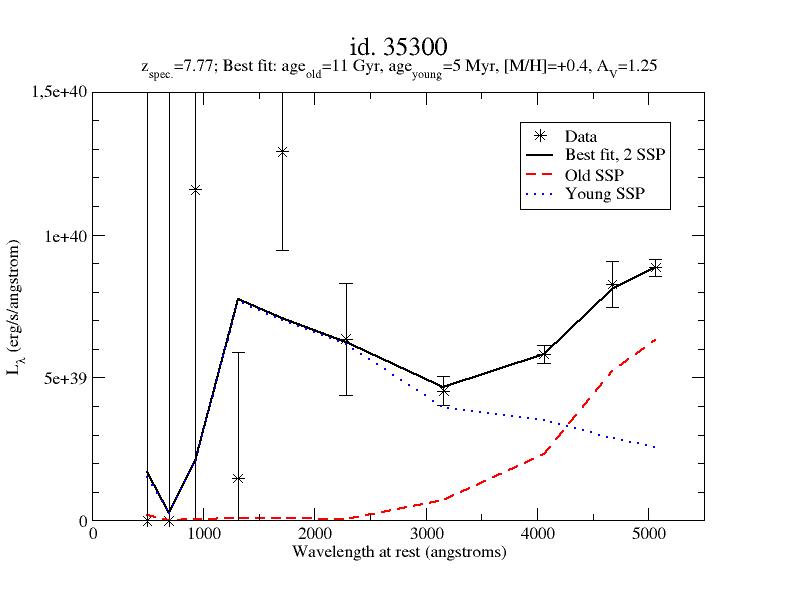}
\vspace{.2cm}
\includegraphics[width=8cm]{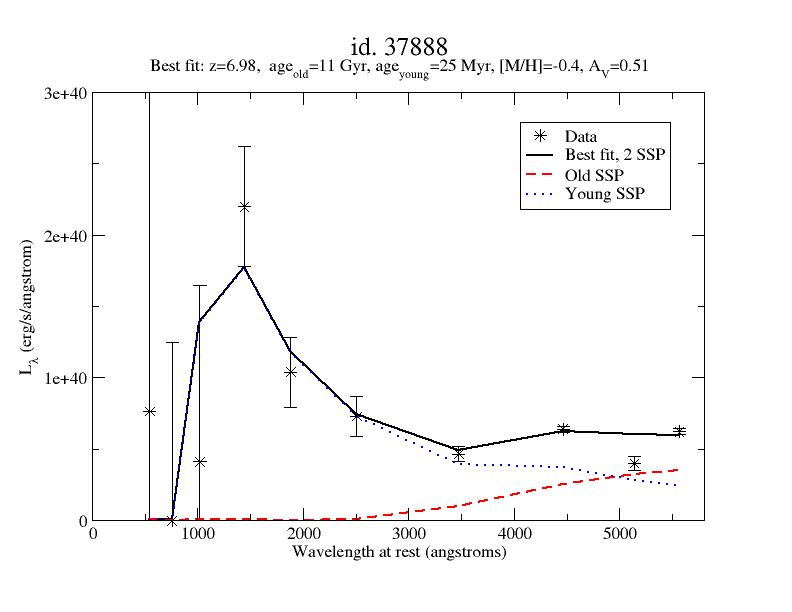}
\hspace{.2cm}
\includegraphics[width=8cm]{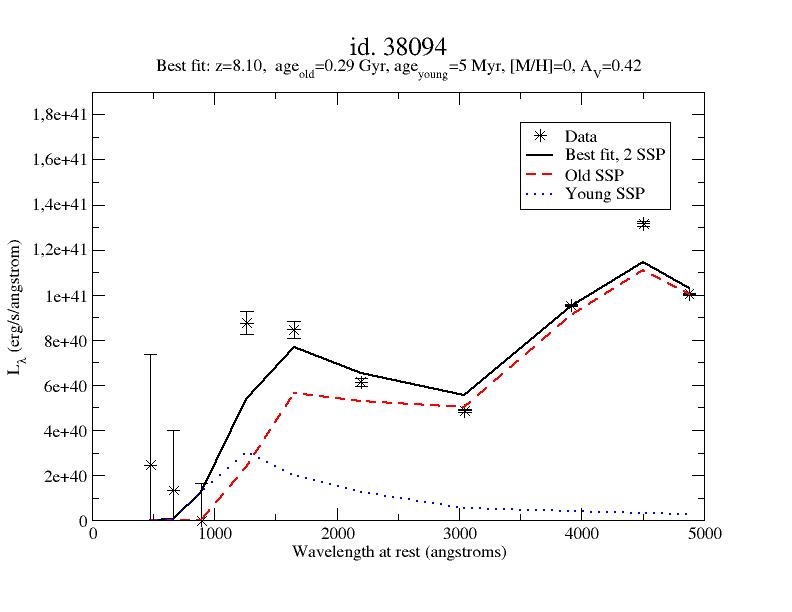}
\vspace{.2cm}
\begin{flushleft}(Fig. \ref{Fig:sedr} cont.)\end{flushleft}
\end{figure*}

\begin{figure}
\includegraphics[width=8cm]{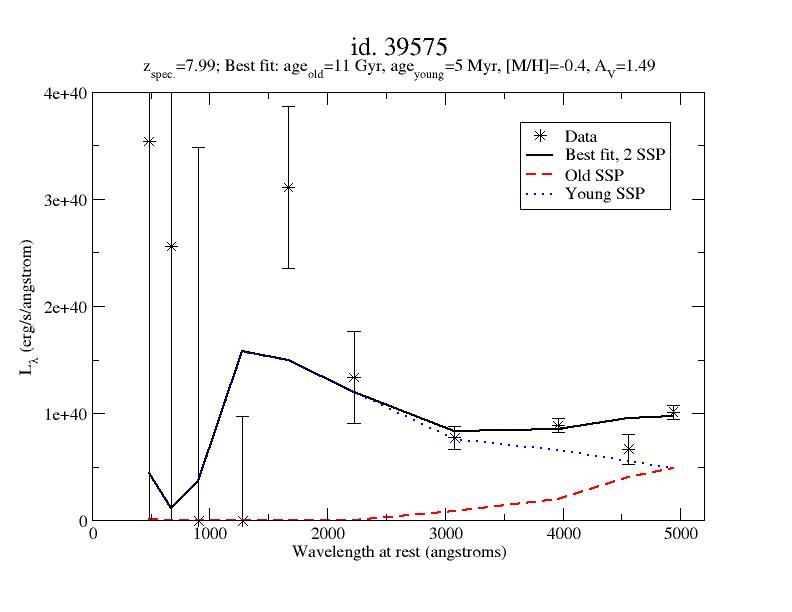}
\vspace{.2cm}
\begin{flushleft}(Fig. \ref{Fig:sedr} cont.)\end{flushleft}
\end{figure}

\subsection{Results}
\label{.results}

The best fit results for the 13 galaxies in our sample are shown in 
Fig.~\ref{Fig:sedr} and Table~\ref{Tab:bestfits}. All cases are compatible 
(within 95\% CL) with zero extinction. The average redshift of these 13 galaxies 
is $\langle z\rangle =8.2\pm 0.4$ (1$\sigma $), at which the $\Lambda$CDM universe 
had an age $t_{\rm Univ}(\langle z\rangle)=0.61\pm 0.04$ (1$\sigma $) Gyr. We 
emphasize that the age of the Universe at any given redshift depends on the 
cosmological model, but the age of the galaxy derived via SED fitting is totally 
independent of the cosmology. 

We emphasize that, for the fits shown in Fig.~\ref{Fig:sedr} and 
summarized in Table~\ref{Tab:bestfits}, we have not introduced any constraint on 
the age of the galaxies, allowing them to be larger than the age of the Universe. 
Were we to limit the age of the galaxies to be younger than that of the Universe, 
the fits would be much worse, as we shall illustrate in appendix \ref{.ageconstraint}.

We estimate the intrinsic (extinction-corrected) luminosity at rest in the 
V-filter, $L_{V,0}$, and the stellar masses of the galaxies corresponding to the SED fitting
assuming for each component (old or young; the young component gives a negligible 
contribution to the mass) a mass-to-light ratio $M/L_{V,rest}\approx 0.6\times 
{\rm age}({\rm Gyr})^{0.7}$ (derived from Fig. 4 of \citet{McL05} for solar metallicity,
neglecting the metallicity dependence). These numbers are given in Table \ref{Tab:masses}. 
The order of magnitude of the stellar masses are similar to those obtained by \citet{Lab23}. 
For the best fit values, we have lower extinction but larger ages than the \citet{Lab23} 
analyses, so these factors largely annul each other, give even larger masses, though we 
have very large error bars in the mass due to the large error bars in the ages. The ratio 
of young/total mass is very low, $\le 1$\% for the best fits, except for the galaxy ID \#11184.

{The weighted average (see appendix \ref{.waver})
of $\log _{10}({\rm age}_{\rm old})$ of these galaxies, converted to linear scale, gives 
\begin{equation}
\label{av1}
\langle {\rm age}_{\rm old}\rangle =1.02^{+0.53}_{-0.37}\;(68.2\% CL)
\ ^{+1.43}_{-0.59}\;(95.4\% CL)\ {\rm Gyr}\,.
\end{equation}

\begin{figure}[h]
\vspace{0cm}
\centering
\includegraphics[width=10cm]{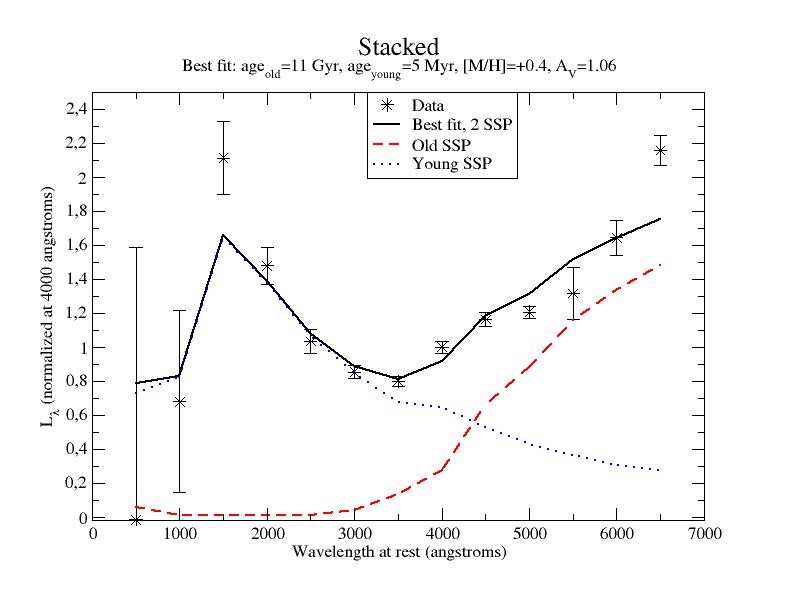}
\vspace{.2cm}
\caption{Best fit using two SSPs of the stacked SED at rest ($\chi ^2_{\rm red}=6.34$).
Data in Table \ref{Tab:stacked}.}
\label{Fig:sedrstacked}
\end{figure}

\begin{table}
\caption{$L_{V,0}$ is the intrinsic (extinction-corrected) luminosity in filter V-rest.
$M$ is the total (young+old) stellar mass. The last column indicates the ratio of young/total mass. 
Error bars correspond to 95\% CL.}
\begin{center}
\begin{tabular}{cccc}
Galaxy ID & $L_{V,0}(L_\odot )$ & $M(M_\odot $) & $M_{\rm young}/M$ 
\\ \hline
2859 &   1.1E10 &  1.2$^{+2.1}_{-0.7}$E10 &  0.03$^{+0.20}_{-0.03}$\% \\
7274 &    2.2E10 &  3.9$^{+2.9}_{-3.8}$E10 &  0.03$^{+0.44}_{-0.03}$\% \\
11184 &   3.8E9 &  8.0$^{+5.4}_{-4.2}$E8 &   19$^{+17}_{-19}$\% \\
13050 &    3.6E10 & 7.7$^{+3.5}_{-7.7}$E10 &  0.2$^{+30}_{-0.2}$\% \\
14924 &    1.8E10 & 5.2$^{+0.3}_{-4.9}$E10 &   0.16$^{+0.49}_{-0.14}$\% \\
16624 &    4.5E10 &  1.3$^{+0.1}_{-1.3}$E11 &  0.04$^{+0.33}_{-0.04}$\% \\
21834 &    3.5E10 &  4.1$^{+0.1}_{-3.5}$E10 &  0.09$^{+0.21}_{-0.08}$\%  \\
25666 &    2.8E9 &  7.5$^{+1.6}_{-7.5}$E9 &   0.3$^{+12}_{-0.3}$\% \\
28984 &   3.5E9 &  9.4$^{+1.2}_{-8.4}$E9 &   0.08$^{+0.64}_{-0.08}$\% \\
35300 &   2.3E10 & 5.9$^{+1.1}_{-5.3}$E10 &  0.12$^{+2}_{-0.12}$\%  \\
37888 &    3.5E9 &  6.7$^{+3.7}_{-6.7}$E9 &   1.0$^{+4}_{-1.0}$\% \\
38094 &  4.3E10 & 1.1$^{+12}_{-0.8}$E10 &    0.13$^{+5}_{-0.13}$\% \\
39575 &   3.6E10 & 7.1$^{+2.7}_{-7.1}$E10 &    0.3$^{+11}_{-0.3}$\% \\ \hline
\end{tabular}
\end{center}
\label{Tab:masses}
\end{table}

\begin{table}
\caption{Stacked SED at rest for the 13 galaxies, and for the 4 galaxies 
with spectroscopic redshift. Fluxes are normalized to unity at 4000 \AA  $N$ is the number of galaxies that have some contribution in the bin;
$N_{\rm sp}$ is the number of galaxies with spectroscopic redshift
that have some contribution in the bin.}
\begin{center}
\begin{tabular}{ccccc}
$\lambda _{\rm rest}$(\AA ) & $N$ & $N_{\rm sp}$ & $\langle F_\lambda \rangle$ (all) & 
$\langle F_\lambda \rangle$ (spec. gal.) \\ \hline
      500 &  13 & 4 & -0.014$\pm $1.599  &  1.726$\pm $4.487 \\
     1000 &  13 & 4 &    0.679$\pm $0.535  & -0.194$\pm $1.541 \\
     1500 &  13 & 4 &    2.112$\pm $0.214  &  2.316$\pm $0.606 \\
     2000 &  13 & 4 &    1.479$\pm $0.110  &  2.144$\pm $0.291 \\
     2500 &  13 & 4 &   1.034$\pm $0.072  &  1.387$\pm $0.199 \\
     3000 &  13 & 4 &    0.853$\pm $0.039  &  0.996$\pm $0.106 \\
     3500 &  13 & 4 &    0.800$\pm $0.030  &  0.917$\pm $0.072 \\
     4000 &  13 & 4 &    1.000$\pm $0.035  &  1.000$\pm $0.049 \\
     4500 &  8 & 4 &    1.163$\pm $0.040  &  1.125$\pm $0.056 \\
     5000 &  6 & 3 &    1.207$\pm $0.035  &  1.529$\pm $0.042 \\
     5500 &  3 & 1 &    1.316$\pm $0.151  &  1.698$\pm $0.063 \\
     6000 &  1 & 1 &    1.643$\pm $0.103  &  1.643$\pm $0.103 \\
     6500 &  1 & 1 &    2.158$\pm $0.088  &  2.158$\pm $0.088 \\ \hline
\end{tabular}
\end{center}
\label{Tab:stacked}
\end{table}

If we sum the 13 SEDs at rest (assuming the redshift inferred from their best fits) 
weighted in proportion to their luminosity at 4000 \AA , we obtain the stacked SED 
plotted in Fig.~\ref{Fig:sedrstacked} and summarized in Table \ref{Tab:stacked}, 
whose optimized parameters (excluding the redshift, which is fixed at zero in the rest SED) are 
given in Table~\ref{Tab:bestfits} (note that the errors in this table are quoted 
as 95\% CL, equivalent to 2$\sigma $). This stacked SED is obtained as a weighted 
(in proportion to the luminosity at 4000 \AA ) sum of SEDs at rest, 
$L(\lambda )=\sum _i L_i(\lambda)$. We include in the calculation of the errors 
the uncertainty in both the flux and the redshift:
\begin{equation}
(\Delta L)(\lambda )=
\end{equation}\[
=\sqrt{\left(\sum _i \Delta L_i(\lambda )\right)^2+\left[\frac{dL(\lambda)}
{d\lambda }\left(\sum _i\frac{L_i(\lambda )}{L(\lambda )}\frac{\lambda 
\Delta z_i}{1+z_i}\right)\right]^2};
\]
$L_i(\lambda )$ and $\Delta L_i(\lambda )$ are derived from the linear 
interpolation of the available values of $L_i(\lambda _j)$ and $\Delta L_i(\lambda _j)$, 
with $\lambda _j$ the wavelengths of the filters at rest. The stacked 
SED is obtained in bins of 500\ \AA , approximately the same as the average 
resolution of each SED at rest.

If we sum the 4 SEDs at rest of only the galaxies with available 
spectroscopic redshift weighted in proportion to their luminosity at 4000 \AA \,
we obtain the stacked SED plotted in Fig.~\ref{Fig:sedrstacked2} and summarized
in Table \ref{Tab:stacked}, whose optimized parameters (excluding 
the redshift, which is fixed at zero in the rest SED) are given in Table~\ref{Tab:bestfits}.

\begin{figure}
\vspace{0cm}
\centering
\includegraphics[width=10cm]{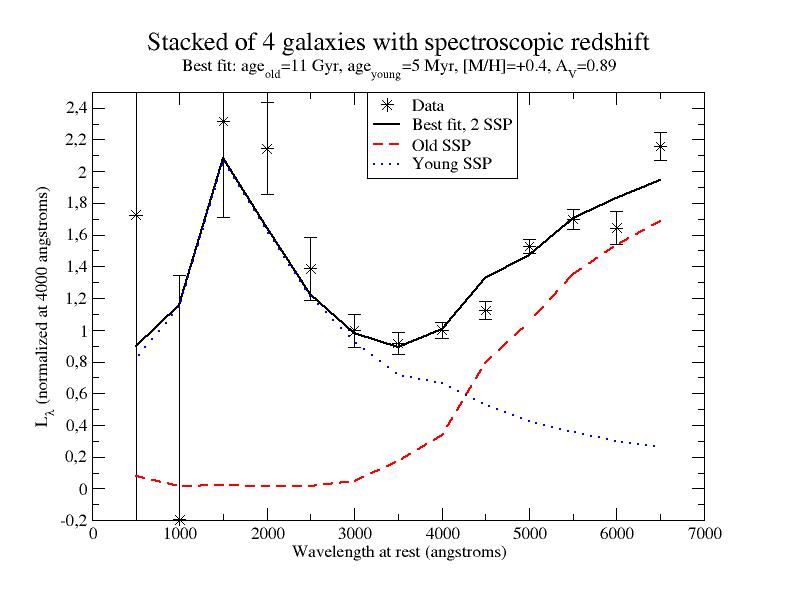}
\vspace{.2cm}
\caption{Best fit using two SSPs of the stacked SED at rest of only the four 
galaxies with spectroscopic redshift ($\chi ^2_{\rm red}=4.08$). Data in Table \ref{Tab:stacked}.}
\label{Fig:sedrstacked2}
\end{figure}

The most interesting parameter of this fit of stacked SED with 13 galaxies is 
the age of the oldest population, which is constrained by Eq.~(\ref{Avni}) to be:
\begin{equation}
\label{av2}
{\rm age}_{{\rm old,\, stacked}}\ge 1.6\ {\rm Gyr} \ \  (68\% CL),\ \ge 0.9\ 
{\rm Gyr} \ \  (95\% CL)\;,
\end{equation}
in agreement with the previous estimate of the weighted average for the minimum 
age. Using other likelihood estimators (see Table~\ref{Tab:maxlh}), the 
probabilities of the young population are even lower. For instance, using a 
maximum likelihood approach (see Appendix \ref{.maxlh}) gives tighter 
constraints: ${\rm age}_{{\rm old,\, stacked}}\ge 3.9\ {\rm Gyr} \ \  
(68\% CL),\ \ge 2.1\ {\rm Gyr} \ \  (95\% CL)$. We keep the result based on
Equation~(\ref{Avni}) since it is more conservative (for it allows a wider 
range of ages). The same exercise stacking only the four galaxies with 
spectroscopic redshifts gives almost the same results: 
${\rm age}_{{\rm old,\, stacked\ spec.}}\ge 1.9$ Gyr (68\% CL),
${\rm age}_{{\rm old, stacked\ spec.}}\ge 0.9$ Gyr (95\% CL).

For $\lambda _{\rm rest}\ge 5500$ \AA , only 1-3 galaxies contributed
to the stacked SED (only 1 in the case of the stacking with spectroscopic redshift). We 
may think that the galaxy or galaxies contributing to $\lambda _{\rm rest}\ge 5500$ \AA (with the
lowest redshifts) are special cases different from the rest of the sample. However,
if we restrict our fits of the SED fitting to $\lambda _{\rm rest}\le 5000$ \AA \ or 
$\lambda _{\rm rest}\le 4000$ \AA , we get very
similar results of the fit (see last rows of Table \ref{Tab:bestfits}).

Note that the errors may be approximately Gaussian for 
$\log _{10}({\rm age}_{\rm old})$, but not for ${\rm age}_{\rm old}$. 
We estimate the probability of fitting the stacked SED of 
all the galaxies with an ${\rm age}_{\rm old}\le 290$ Myr to be
$3\times 10^{-4}$, (derived from the $\chi ^2_{\rm red}=31.25$ of the 
best fit using ${\rm age}_{\rm old}\le 290$ Myr; much larger than the 
corresponding $\chi ^2_{\rm red}=6.34$ for the best fit using 
${\rm age}_{\rm old}\le 11$ Gyr). This would be equivalent to a 
3.6$\sigma $ event. Based on a maximum likelihood approach 
(Appendix~\ref{.maxlh}), it would reach 4.9$\sigma $. The probability 
of fitting the stacked SED with an ${\rm age}_{\rm old}\le 640$ Myr is
$1.8\times 10^{-3}$, derived from the $\chi ^2_{\rm red}= 27.19$ of the 
corresponding best fit (see the left panel of Fig.~\ref{Fig:sedrstacked3}),
which would be equivalent to a 3.1$\sigma $ event in the most 
conservative calculation. It is 2.4$\sigma$ for the stacked SED with 
four spectroscopic galaxies.

\begin{figure}
\vspace{0cm}
\centering
\includegraphics[width=8cm]{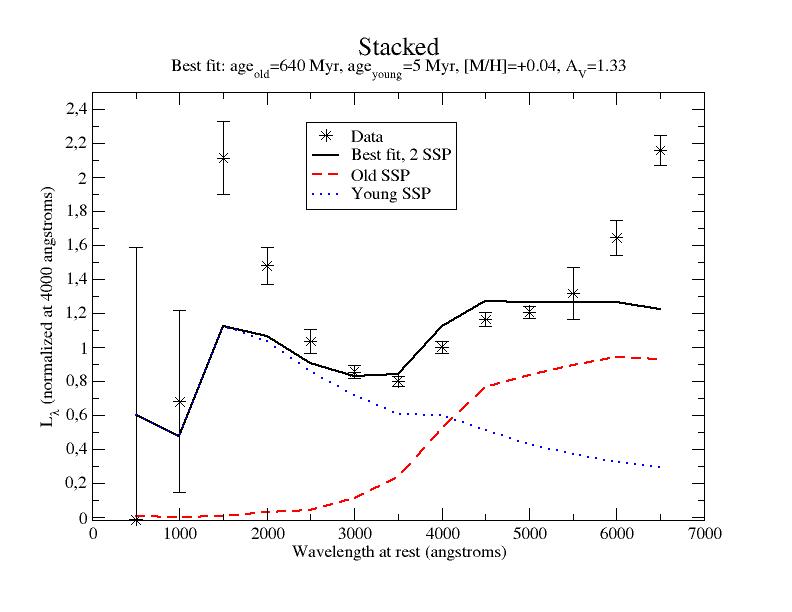}
\vspace{.2cm}
\includegraphics[width=8cm]{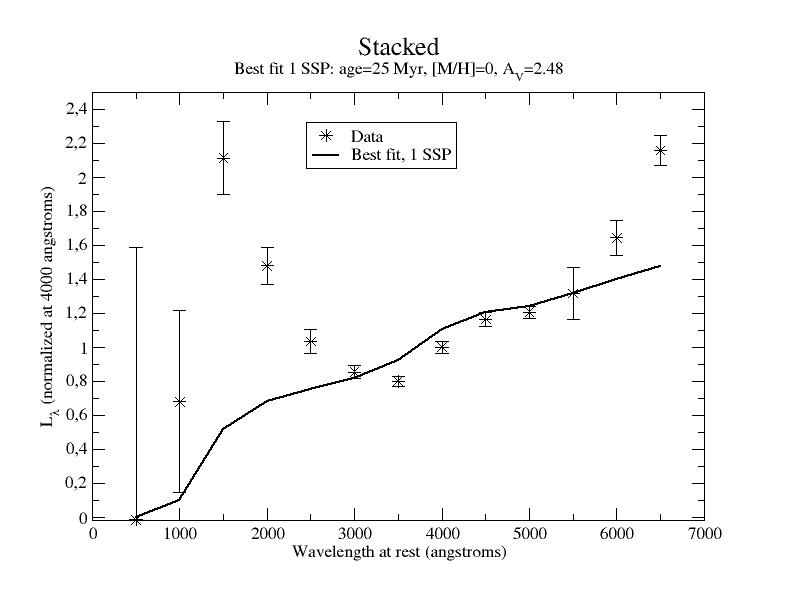}
\vspace{.2cm}
\includegraphics[width=8cm]{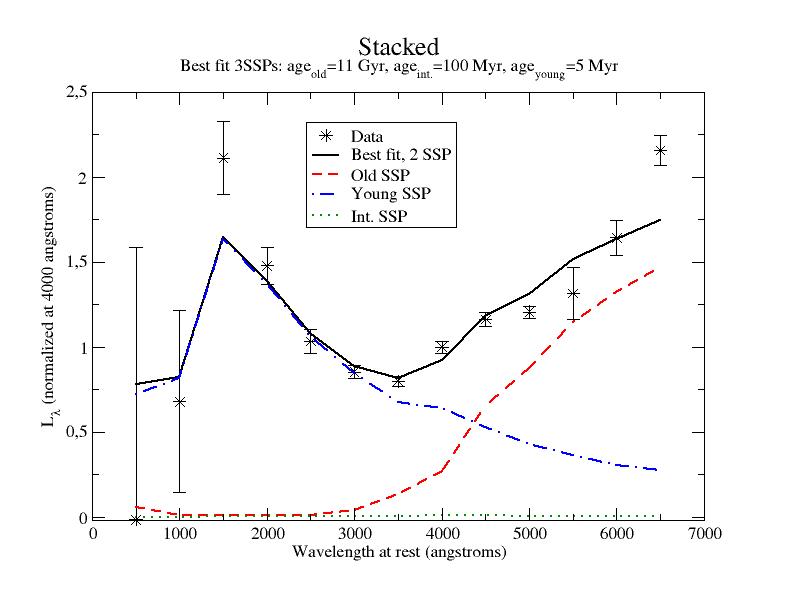}
\caption{Best fit of the stacked SED at rest: 
Left panel: two SSPs with age$_{\rm old}\le 640$ Myr 
($\chi ^2_{\rm red}=27.19$); Middle panel: one SSP
($\chi ^2_{\rm red}=24.30$); Right panel: three SSPs 
assuming $[M/H]=+0.4$, $A_V=1.06$ ($\chi ^2_{\rm red}=6.36$).}
\label{Fig:sedrstacked3}
\end{figure}

A fit of this stacked SED with one SSP would yield much worse results 
in comparison with two SSPs, as may be seen in the middle panel of
Fig.~\ref{Fig:sedrstacked3}. One cannot fit both breaks at 
the same time with only one population. Increasing the number of SSPs 
beyond two does not improve the fit significantly. For instance, an 
optimized SED with three SSPs (while keeping the extinction and metallicity 
at their best fit values for two SSPs, in order to keep the same number of 
free parameters) produces almost the same best-fit value of the oldest 
population and its lower age limit, shown in the right panel of Fig.~\ref{Fig:sedrstacked3}.
Evidently, the third age produces only a very small additional 
contribution to the young population. This reinforces our previous claim 
that exponentially decaying, extended star formation models do not fit these 
V-shaped SEDs. Nonetheless, a more accurate determination of extinction models 
and templates of SSPs, and with more precise values of ages and metallicities
instead of the discrete values we have used, might improve the fits.

To check the consistency of our calculations with those based on 
publically-available SED fitting packages, we compare in Appendix~\ref{.lephare}
our results with those produced by {\it le PHARE}. These yield similar conclusions 
regarding the necessity of including an old population to fit the data.

Our assumptions may harbor other systematic errors: neglecting the light 
contamination from other galaxies or the assumption of a standard IMF, and 
the presumed extinction law. The super-solar metallicities in 
the GALAXEV model are produced by interpolating the grids, which might yield
some systematic error in the star-formation histories and hence the ages.
From our analysis, however, we can see that the metallicity is not a significant 
factor in the SED fitting. But these possible uncertainties should not 
prevent us from seeing the fundamental outcome: it is not possible to fit 
the observed SED with a pure young population because of the clear existence 
of the V-shape (with pronounced wings) and a Balmer break \citep{Lop17} and 
red color at longer wavelengths \citep{Lop10}. 

In addition, the amplitude of the Balmer break, which is strongly dependent 
on age \citep{Ste23}, necessarily should have $\sim 1$ Gyr-old populations, 
though some authors believe that the JWST high-z galaxies are compatible with
the prediction of simulations using the standard cosmological model \citep{Vik24}. 
The possibility of explaining the Balmer break in double-break galaxies with 
emission lines \citep{Des23} is already contained in our fits because the
\citet{Bru03} stellar population models contain the possibility of including 
any fraction of a young population, which includes emission lines.

Other results obtained with similar methods of SED fitting by \citet{Roc18} 
also yield quite high ages at very high redshifts: a 450-500 Myr-old galaxy 
at $z=8.6$ and a $\sim 160$ Myr-old galaxy at $z=11.1$.

\subsection{Understanding the results}

One may suspect that the numerical SED fitting may have errors that produce the 
result that the galaxy must be old, even if our analysis includes an independent 
check using different software (Appendix \ref{.lephare}). When using a code as a 
black box that receives some input and delivers some output, strange results can
arise if the algorithm has errors. It is therefore useful to understand and
interpret the physical meaning of any numerical result and the reason it is 
derived. As we have already noted on several occasions, the key issue to 
understand with our analysis is that the large ages we infer for the old stellar 
population are constrained by the Lyman break to have small extinction, so reddening 
can affect the Balmer break only by a small amount as well. But we have red color
around the Balmer break which, in the absence of large extinction, can thus be 
due solely to an old population. Let us confirm this with a rough calculation of 
the V-shaped SED.

The first element is that the far-UV can only be explained with a very young 
population; an old population makes only a negligible contribution there. If we take 
our stacked spectrum of 13 galaxies in Fig.~\ref{Fig:sedrstacked} (as we have seen 
previouly, there is no reason to suspect that errors in the photometric redshift 
produce important artefacts in the stacked SED at rest, because we get the same SED
using only the 4 galaxies with spectroscopic redshift), we measure 
$\frac{F_{\lambda }(1500\ \AA)}{F_{\lambda }(3000\ \AA)}=2.48\pm 0.23$. Hence, 
$m_{\rm AB,rest:1500-3000\ \AA}=0.52\pm 0.10$. This is a relatively blue color in 
the far-UV and is expected because the selection criteria used to choose these 13 
galaxies include the constraint F150W-F277W$<0.7$ which, at an average redshift of 
$z=8.2$, is equivalent to setting $m_{\rm AB,rest:1630-2980\ \AA}<0.7$. 

Let us ignore the dependence on metallicity, which is low as we have seen, and 
also neglect any possible AGN contamination which, as we shall confirm in the next
subsection, is also low. We consider only solar metallicity here. Then, assuming 
an age of 5 Myr for the young population to fit the far-UV signal, for which 
$m_{\rm AB,rest:1500-3000\ \AA}$[SSP:5 Myr]$=-0.34$ and given the reddening 
implicit in Calzetti's law, $E({\rm rest}:1500-3000\ \AA)=0.72A_V$, we get 
$A_V=\frac{0.52-(-0.34)}{0.72}=1.19$ (quite close to the result of our SED 
fitting corresponding to Fig. \ref{Fig:sedrstacked}). This is the maximum 
extinction, because for a larger age of the young population, the intrinsic 
color is redder, thus yielding a smaller $A_V$. Therefore, it makes sense to 
claim that our average $A_V$ is, at most, one magnitude, for otherwise the 
far-UV region would be redder and the Lyman-break might even vanish.

How big should our young population be to fit this Lyman break? We know that 
at 1500 \AA \ at rest the contribution from the old population is negligible
and, comparing it with the relative flux at the V-rest filter, we derive 
(using Eq.~\ref{Ftheor}, bearing in mind the normalization $L_{\rm SSP} [\lambda =
5500\ \AA \equiv V]=1$, and adding Calzetti's law extinction),
\begin{equation}
\frac{L_{\lambda }(1500\ \AA)}{L_{\lambda ,V}}=\frac{A_2L_{\lambda, 
{\rm young}}(1500\ \AA)}{1+A_2}10^{-1.55A_V/2.5},
\end{equation}
so that 
\begin{equation}
\label{a2}
A_2=\left[\frac{L_{\lambda ,{\rm young}}(1500\ \AA)\times 10^{-1.55A_V/2.5}}
{\frac{L_{\lambda }(1500\ \AA)}{L_{\lambda ,V}}}-1\right]^{-1}.
\end{equation}
In our stacked SED, we measure $\frac{L_{\lambda }(1500\ \AA)}{L_{\lambda ,V}}=1.60\pm 0.25$.

Next, we analyze the redder wavelengths and consider the color $(B-V)_{\rm AB}$. For the 
combination of two SSPs we have adopted, we get
\begin{equation}
\label{bmv}
(B-V)_{\rm AB}=(B-V)_{\rm AB,old}+E(B-V)
\end{equation}\[\ \ \ \ \ \ \ \ \ \ 
-2.5\log_{10}\left(\frac{1+A_2\frac{L_{\lambda, 
{\rm B,young}}}{L_{\lambda ,{\rm B,old}}}}{1+A_2}\right).
\]
The reddening with Calzetti's law is $E(B-V)=0.247A_V$. The third term on the righthand 
side gives the blueing due to the young population's contribution, and depends on the 
calculated fractions of young and old components. The young population is always bluer 
than the old population, so this term is always negative. The value of this third term 
falls within the limits $(-0.54,-0.18)$, consistent with range of ages compatible with
the global color, as we shall see in the next paragraph.

In the stacked SED, we measure $\frac{F_{\lambda }(4400\ \AA)}{F_{\lambda } (5500\ \AA)}=
0.78\pm 0.04$, for which $(B-V)_{\rm AB}=0.74\pm 0.06$. Again, this is expected because we 
have preselected red galaxies through the condition F277W-F444W$>1.0$, equivalent (at average 
redshift $z=8.2$) to $m_{\rm AB,rest:2980-4830\ \AA}>1.0$. If one were to include only
extinction-correction and ignore the blueing of the young population, the result would
be $(B-V)_{\rm AB,old}>0.45\pm 0.06$, equivalent in Vega calibration to $(B-V)_{\rm Vega,old}>0.57 
\pm 0.06$. The blueing of the young stellar population would make the old population even 
redder, with a larger $(B-V)_{\rm Vega,old}$ and never smaller. This color is associated 
with a stellar population older than 1 Gyr \citep[Fig. 2]{Lop10}. A population younger than 
0.5 Gyr would require $(B-V)_{\rm Vega,old}<0.35$, which cannot be obtained with these data.
The fact that our preferred solution (though with a very generous error bar) includes 
the oldest population is due to the fact that the data favor the value $(B-V)_{\rm Vega,old}
\approx 1.1$, for which the third term in Eq.~(\ref{bmv}) contributes with a value of -0.5, 
though a broad range, $(B-V)_{\rm Vega,old}>0.57$, is possible.

In other words, in order to make the observed red color after the Balmer break compatible
with an age of the oldest population lower than 0.5 Gyr, we would need to make 
$(B-V)_{\rm Vega,old}<0.35$ and an extinction $A_V\gtrsim 2.1$, which would make 
$E({\rm rest}:1500-3000\ \AA)\gtrsim 1.5$, $m_{\rm AB,rest:1500-3000\ \AA}\gtrsim 1.2$, 
so we would not see the Lyman break. Of course, this is forbidden by our data that 
a priori have a selection with $m_{\rm AB,rest:1630-2980\ \AA}<0.7$.

\subsection{AGN contamination}
\label{.AGN}
There are several color criteria for selecting AGNs in high redshift samples: 
some of them use infrared mid-infrared colors \citep{Ass10}, or are also 
especially designed for JWST near-infrared filters \citep{Gou22}. These 
criteria are not applicable to our particular selection of galaxies, 
however, because: (i) we do not have photometry beyond 5$\mu $m; (ii) our 
galaxies were selected to be very red and provide an important Balmer break, 
so we cannot use the filters F356W, F444W; and (iii) though the filters 
F150W, F200W, F277W might be used as criteria for the JWST filters 
\citep{Gou22}, this is only practical when the signal/noise is relatively 
high, which is not our case here.

An important AGN component could modify the fitted parameters 
and also the mass of the galaxy, but it is not expected in our sample.  AGNs 
are extremely blue in the range below $\sim 1\,000$ \AA \ at rest, unless 
there is a huge extinction, which is not expected because otherwise we could 
not see the Lyman break. Strong emission lines, a dusty continuum 
and AGN contributions together could explain this Lyman break \citep{Win23}, 
but our model contains all of these elements and does not favor this solution, 
as we shall see below. We again note that emission lines are already included
in the young population templates in the GALAXEV model, and lines are also
included in the AGN template we shall use.

The sample of 13 galaxies selected by \citet{Lab23} already removed most of 
the strongly dominant AGNs by requiring that they were not detected in the HST 
filters [SNR(F435W, F606W, F814N)$<2$] (equivalent to $\lambda \lsim 1\,000$ \AA 
\ at rest for an average redshift of $z=8$). There may be still be some possible 
small component of AGN or active galaxy with small black hole mass, however. 
As a matter of fact, one of those galaxies (id. 13050) has already been identified 
as an AGN with a low-mass ($\sim 10^7$ M$_\odot $) black hole at $z=5.62$ via a 
follow-up spectroscopic observation \citep{Koc23}. The noise in HST filters is 
even higher than in JWST filters, so some emission at $\lambda \lsim 1\,000$ \AA 
\ at rest may be hidden. An AGN component would make a blue contribution around 
the Balmer break, so it cannot explain the large Balmer break we observe here, 
and the higher the AGN component, the larger the age of the oldest population 
necessary to redden the Balmer break (apart from extinction), so it will not 
affect our calculations of the minimum age of the oldest population. 
Nevertheless, we shall carry out a fit of the stacked SED 
including an AGN component to check that it does not affect our results.

In this additional fit, we assume the usual two SSPs of stellar population 
plus an extra AGN component with fixed characteristics (so there are no extra 
free parameters):
\begin{equation}
F_{\rm theor.}(\lambda _i)=\frac{L_0}{4\pi d_L(z)^2(1+z)}
\end{equation}\[
\times
[\langle L_{\rm SSP}
({\rm age _{\rm old},[M/H]}, A_V;\lambda /(1+z))\rangle _T
\]\[
+A_2\langle L_{\rm SSP}
({\rm age _{\rm young},[M/H]}, A_V;\lambda /(1+z))\rangle _T
\]\[
+A_3\langle L_{\rm AGN}(A_{V,AGN};\lambda /(1+z))\rangle _T]
\,,\]
where $L_{\rm AGN}$ is obtained from \citet{Ass10}, normalized 
such that $L_{\rm AGN}(A_V; 5500\ \AA)=1$, and $A_3$ represents the ratio of 
AGN/old rest luminosity at 5500 \AA \ when the extinction for AGN and stars
is the same. Fig.~\ref{Fig:sedrstacked4} shows the best fit of the stacked 
SED at rest for $A_3=0.0045, 0.023, 0.090$ including the same extinction
that affects the stellar populations ($A_{V,AGN}=A_V$), characterized by 
$\chi ^2_{\rm red}=$ 6.41, 7.01, 10.98 respectively, and with an age of the 
oldest population equal to 11 Gyr in all cases. Fig.~\ref{Fig:sedrstacked4ne} 
shows the best fit of the stacked SED at rest for $A_3=0.0045, 0.023, 0.090$ 
without any extinction for the AGN component, characterized by 
$\chi ^2_{\rm red}=$ 7.87, 21.04, 52.86 respectively, and with an age of the 
oldest population equal to $5-11$ Gyr. Fig.~\ref{Fig:sedrstacked4ee} shows the 
best fit of the stacked SED at rest for $A_3=0.127, 0.575, 5.81$ with a heavy 
extinction for the AGN component, $A_{V,AGN}=3$, representing a red QSO 
\citep{Cal21}, characterized by $\chi ^2_{\rm red}=$ 6.65, 11.28, 26.25 
respectively, and with an age of the oldest population equal to 11 Gyr. 

A usual blue AGN with small extinction component makes the 
flux bluer around the Balmer break, which requires an even greater
age for the galaxy, or increasing the reddening of the stellar population 
through extinction, but this is limited due to the presence of the Far-UV 
features, especially the bin of 500 \AA \ at rest (equivalent to 
the observed flux in the filter HST/F435W).

A red/heavily-extincted AGN may help to explain the red color near the
Balmer break, but its amplitude would also be limited by the flux at 
500 \AA \ at rest, which not even heavy extinction can remove. This is 
due to the strong decrease in extinction below 800 \AA \ and a heavy 
increase in the intrinsic luminosity at these wavelengths: 
$L_{\rm AGN}(A_V=0;500\ \AA )=260\,L_{\rm AGN}(A_V=0;5500\ \AA )$, $A_{500 \AA}=2.5\,A_V$,
whereas $L_{\rm AGN}(A_V=0;800\ \AA )=95\,L_{\rm AGN}(A_V=0;5500\ \AA )$, $A_{800 \AA}=5.1\,A_V$.
Thus, even with a heavy extinction of $A_V=3$, we get $L_{\rm AGN}(A_V=0;500\ \AA )
\sim 4\,L_{\rm AGN}(A_V=0;5500\ \AA )$, even higher when we take an average over 
the wavelengths within the wide filter. Moreover, a heavily extincted QSO dominating 
the flux contribution would need a very large intrinsic luminosity of the QSO (to 
compensate for the extinction). In our example with the largest QSO contribution 
and largest extinction (Fig. \ref{Fig:sedrstacked4ee} with $A_3=5.82$), we would 
need an intrinsic QSO luminosity $L_{0;300-6500\ \AA}\sim 10^{47}$ erg/s , which 
requires a massive black hole whose formation would conflict with the short
timeline available at $z\approx 8$ within standard cosmology. This is not 
necessarily impossible, since something similar has already been observed 
\citep{Wan21}, but not with such an exotic combination of extinction plus 
an extinction-free galaxy.

Summing up, an AGN component does not help to solve the tension, but actually 
makes it even worse.

\begin{figure}
\vspace{0cm}
\centering
\includegraphics[width=8cm]{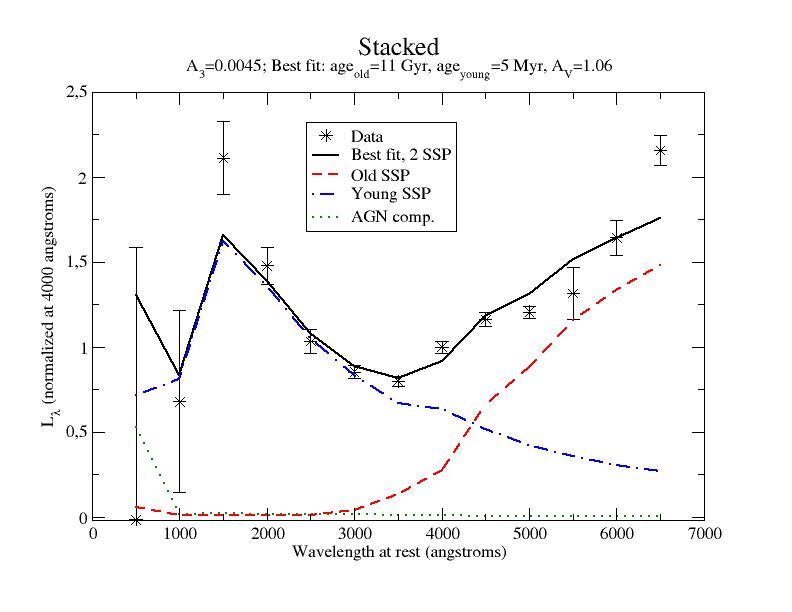}
\vspace{.2cm}
\includegraphics[width=8cm]{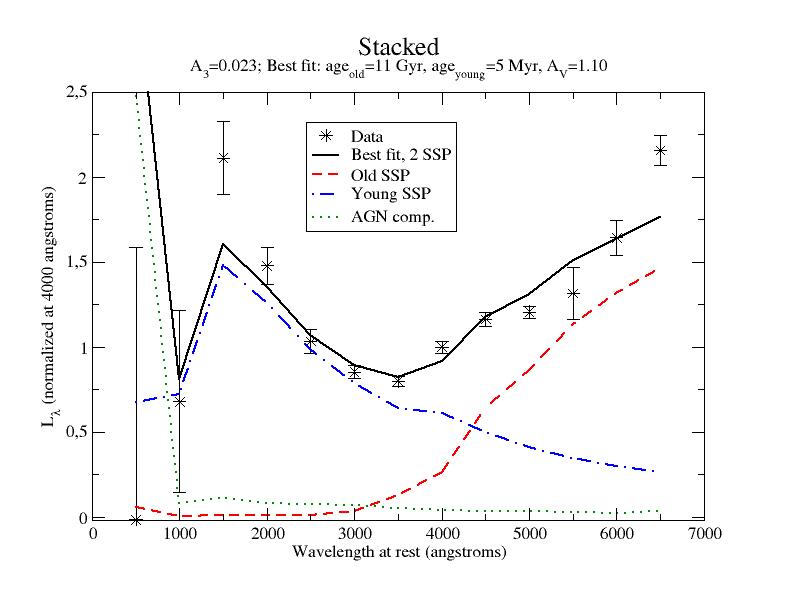}
\vspace{.2cm}
\includegraphics[width=8cm]{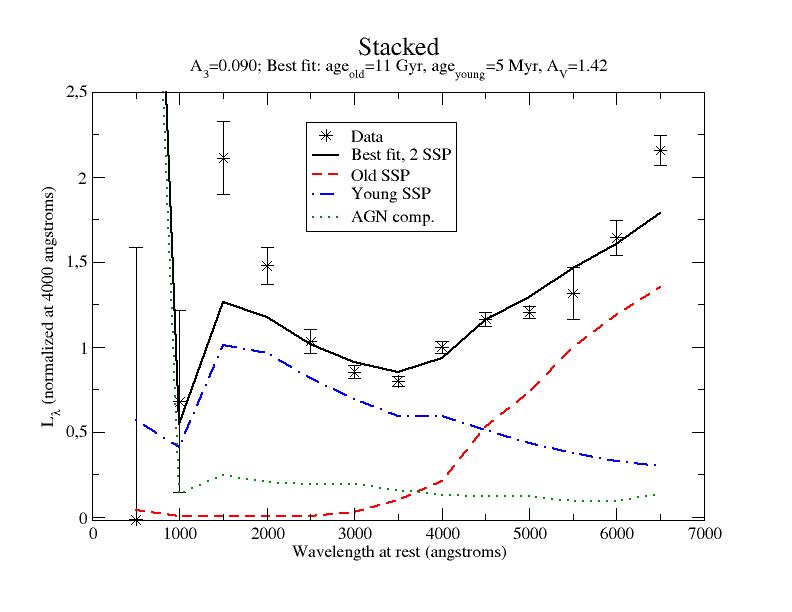}
\caption{Best fit of the stacked SED at rest including an AGN 
component with $A_3=0.0045, 0.023, 0.090$, respectively, and the same extinction 
($A_{V,AGN}=A_V$) that affects the stellar populations.}
\label{Fig:sedrstacked4}
\end{figure}

\begin{figure}
\vspace{0cm}
\centering
\includegraphics[width=8cm]{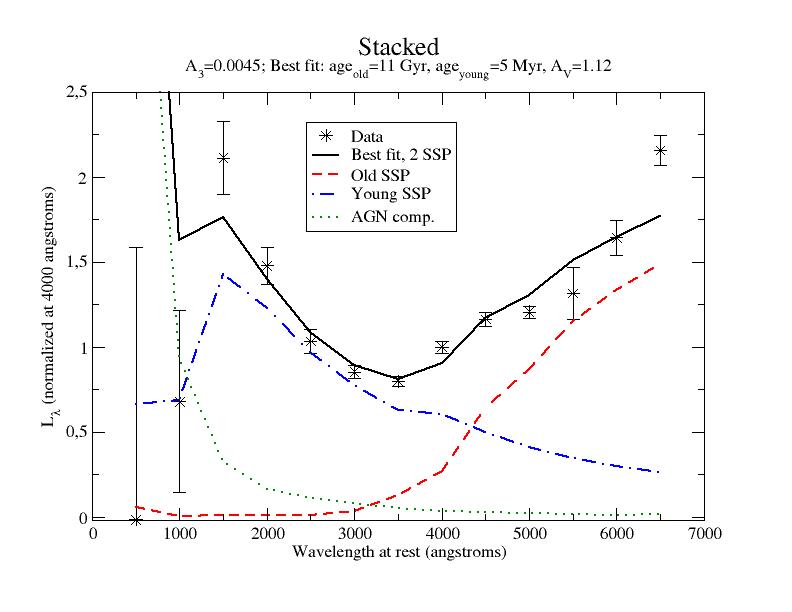}
\vspace{.2cm}
\includegraphics[width=8cm]{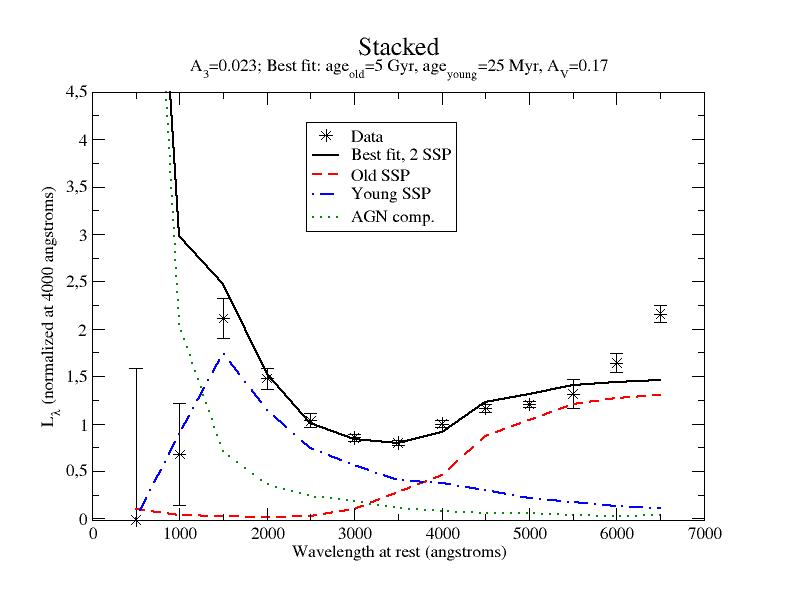}
\vspace{.2cm}
\includegraphics[width=8cm]{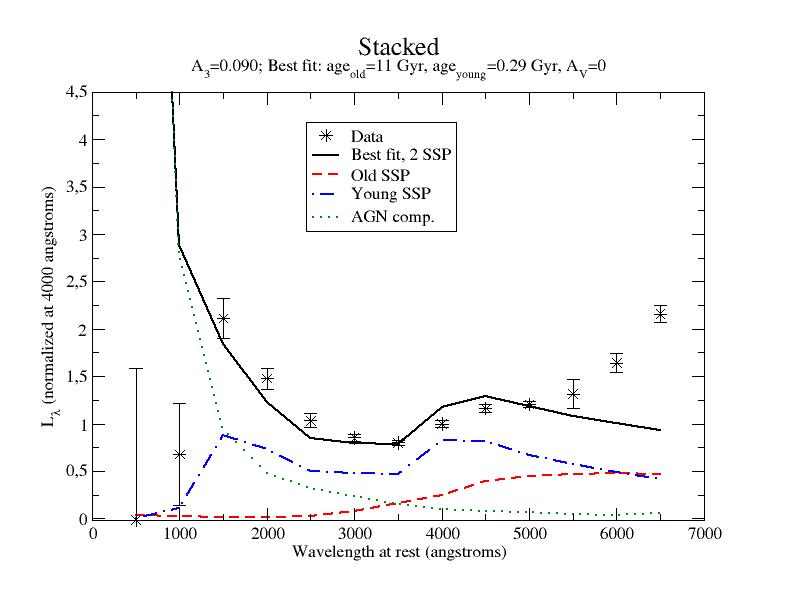}
\caption{Best fit of the stacked SED at rest including an AGN 
component without extinction, $A_{V,AGN}=0$, and with $A_3=0.0045, 0.023, 0.090$, respectively.}
\label{Fig:sedrstacked4ne}
\end{figure}

\begin{figure}
\vspace{0cm}
\centering
\includegraphics[width=8cm]{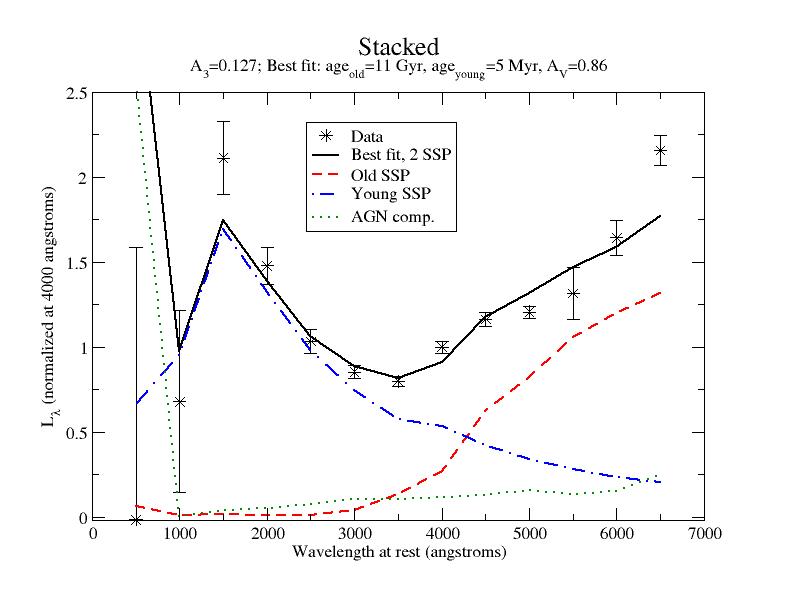}
\vspace{.2cm}
\includegraphics[width=8cm]{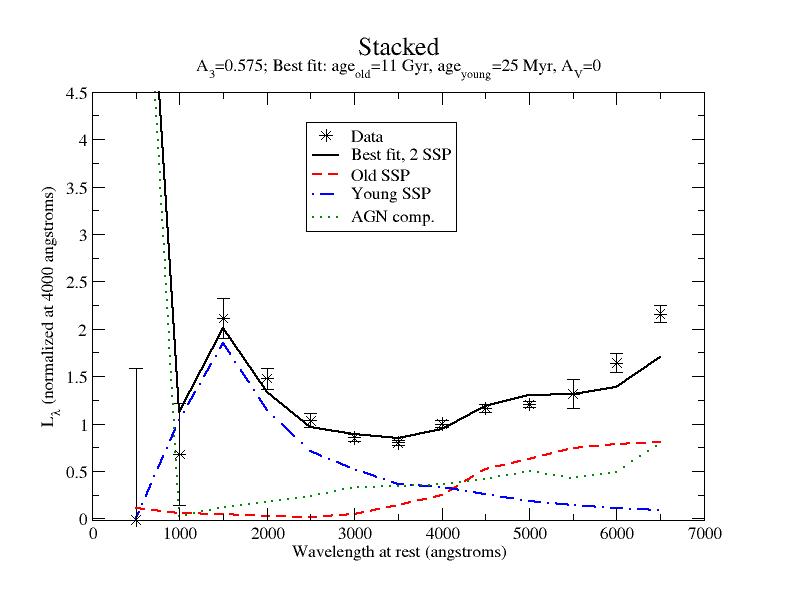}
\vspace{.2cm}
\includegraphics[width=8cm]{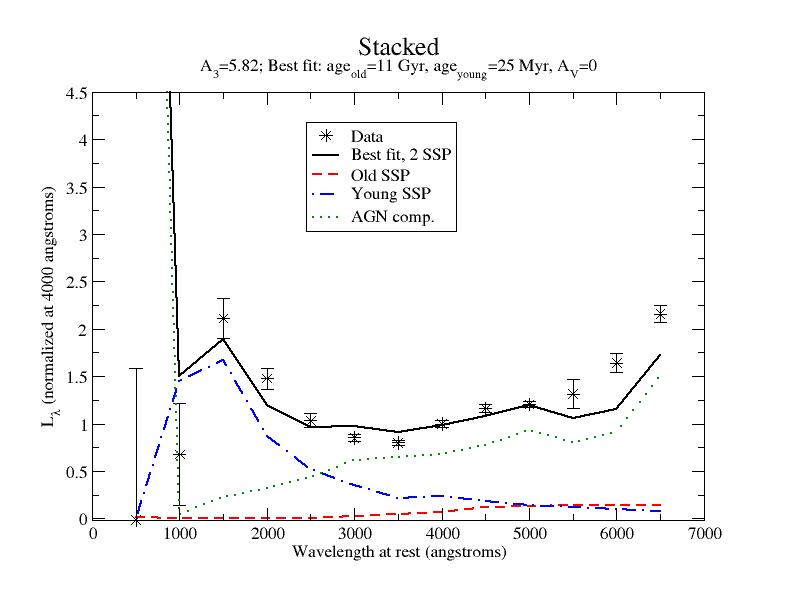}
\caption{Best fit of the stacked SED at rest including an AGN 
component with extinction, $A_{V,AGN}=3$, and $A_3=0.127, 0.575, 5.82$, respectively.}
\label{Fig:sedrstacked4ee}
\end{figure}

Other analyses \citep{Koc23,Bar24} seem to suggest solutions involving 
AGN components with no extinction near the Lyman-break and heavy extinction near 
the Balmer break. This hybrid incorporates a young galaxy free of extinction
and a heavily obscured and red QSO, or a blue QSO plus a dusty galaxy and 
a pure QSO plus a torus or similar exotic multicomponent galaxies that have 
never been observed in a single object at low $z$. 
 
Certainly, if we have more free parameters in a model than the number of data, we 
might be able to fit anything, but it is not clear whether these solutions represent 
real single galaxies. In any case, even these exotic configurations are confronted
with problems: a blue/unextincted QSO is excluded by the non-detection of the galaxies 
in the filter HST/F435W (which is not used in \citet{Koc23,Bar24}); a red QSO might 
be produced by heavy extinction in the nuclear volume \citep{Cal21} but, as we have seen,
this would also leave a very significant imprint in the filter HST/F435W. The 
non-detection in the filter HST/F435W is the key to excluding the strong AGN 
contribution hypothesis.

\section{Discussion}
Our overall analysis, combining Equations~(\ref{av1}) and (\ref{av2}), 
yields a galaxy age, age$_{\rm old}$, between 0.9 and 2.4 Gyr at 95.4\% CL, and
between 0.65 and 3.80 at 99.7\% CL. Galaxies within $2<z<4$ with V-shaped SED and
similar colors at rest have also been analyzed with the same technique, and
it was concluded that for the galaxies with $z > 2.5$ there is no
significant evolution of their average age, with all average ages of galaxies 
mostly remaining between 1 and 2 Gyr \citep{Gao24}. Apparently, JWST has observed 
at $z=8$ the same type of galaxies and with the same ages than those ones analyzed 
with ZFOURGE data by \citet{Gao24} within $2.5<z<4.0$.

The new JWST observations have thus uncovered,
not only many galaxy candidates with photometric redshifts larger than 10 
\citep[e.g.,][]{Mel23,Fra23,Whi23,Gla23,Cur23}, but also the very massive 
galaxies we have analyzed in this paper that are older than the Universe in 
the context of $\Lambda $CDM within $\approx 95$\% CL or, in any case, 
within $>99.9$\% CL, formed well beyond $z=10$. This more robust sample of 
structures provides even tighter constraints on galaxy formation theory in 
$\Lambda $CDM, particularly from the more compelling estimate of their 
photometric redshifts based on the two breaks in their SED. These results 
conflict with semi-analytical $\Lambda $CDM models claiming that very massive 
galaxies formed much later \citep{Guo11}.

The principal difficulty facing the standard model in accounting for the 
very early emergence of galaxies has typically been addressed by enhancing 
star formation, the growth of halos, stochasticity or evolving initial mass 
functions in the early Universe \citep{Fin23}. But the JWST observations
have pushed the threshhold for these processes too uncomfortably close to the big 
bang itself. We appear to have exhausted any flexibility left in understanding how
the primordial plasma could have cooled and condensed rapidly enough to form these
structures by $\sim 300$ Myr in the context of $\Lambda$CDM \citep{Mel23}. Without
much doubt, some significant new physics would be required to adequately explain
how such massive structures could have formed so quickly in the $\Lambda$CDM
Universe.

Modelling the extinction of high-$z$ galaxies with a dust similar 
to local galaxies may require a reexamination \citep{Mar23}, though the effects 
are expected to be small since all our fits are compatible with no extinction; 
indeed the presence of Lyman-$\alpha $ breaks at far-UV indicates that we should 
have very little amount of dust, whatever their properties are. In any case, it 
is important to remark that using mathematical fitting models with much lower
amounts of dust around young populations than is assumed for the old populations 
is unphysical, and here we have avoided it.

Stellar population synthesis is based on models, but we must also bear in mind 
that the assumed cosmology is itself only a working model, and should be viewed 
as being just as adjustable as stellar formation theory.  Though one 
may question the validity of the GALAXEV templates to accurately represent the 
stellar populations of a given age and other assumptions, one can argue that 
cosmology itself is no more robust than stellar astrophysics. Assuming that 
our calculations are correct, we do not believe that it is far more likely 
for the stellar astrophysics and the other assumptions and approximations 
we have used throughout this paper to be flawed rather than the cosmology. 
In order to know whether one or the other is more likely, we would need to 
calibrate their probabilities based on sound statistics. For instance,
what is the probability that the GALAXEV models (or any other Stellar 
Population Synthesis model, such as the GISELL library with a Salpeter IMF 
that was also used in the appendix \ref{.lephare}) are incorrect? 
Thus far, these stellar synthesis models have worked well in fitting observations
of local galaxies and are consistent with our current knowledge of stellar 
astrophysics. Though these approximations may deviate significantly from reality,
we cannot calculate how likely that is. 

The standard model of cosmology accounts rather well for many observations, 
but there are also many instances that create (sometimes significant) tension 
with the predictions of $\Lambda $CDM \citep[e.g.,][]{Lop22}. The probability 
of $\Lambda $CDM being correct can only be determined a posteriori, once we 
definitively solve the puzzle of the Universe. In the end, astrophysics is a 
science largely based on 'orders of magnitude', not the fine precision of
some theories tested in laboratories on Earth. The work we have reported
here relies on several approximations that need to be tested further before 
we can conclude that standard cosmology is wrong.

Of course, not all 
alternatives to $\Lambda$CDM fare better. For example, were we to consider a 
cosmology without dark energy, such as Einstein-de Sitter, the problem would 
be worse. In this cosmology, $\langle z\rangle =8.2\pm 0.4$ (1$\sigma $)
would correspond to an age $t_{\rm Univ}(\langle z\rangle)=0.33\pm 0.02$ 
(1$\sigma $) Gyr, which would also lead to a nonsensical time of birth for
the JWST galaxies prior to the big bang within a $>99$\% CL.

Other cosmological scenarios might account for these observations better than 
$\Lambda$CDM, however. For example, the $R_{\rm h}=ct$ universe 
\citep{MeliaShevchuk:2012,Melia:2020,Mel23} gives an age of the Universe of 
14.6 Gyr, or Covarying Coupling Constants+ Tired-light (CCC+TL) hybrid model 
\citep{Gup23} gives an age of the Universe of 26.7 Gyr (though difficult to 
reconcile with the age of the oldest known sources, in our opinion). In the 
most reasonable alternative, i.e., the $R_{\rm h}=ct$ universe, the timeline 
at $z\gsim 6$ is approximately twice as long as that in the standard model,
assuming $H_0=70$ km s$^{-1}$ Mpc$^{-1}$. In this cosmology, 
$\langle z\rangle =8.2\pm 0.4$ (1$\sigma $) corresponds to an age 
$t_{\rm Univ}(\langle z\rangle)=1.51\pm 0.07$ (1$\sigma $) Gyr, which leads 
to $z_{\rm form}\gtrsim 21$ (95.4\% CL) ($t_{\rm Univ}(z_{\rm form})<0.61$ Gyr). 
This provides a suitable scenario where these massive galaxies started forming 
enough time after the Pop III stars are believed to have first appeared. 
Together with also providing an explanation for how the earliest (though less 
massive) galaxies would have emerged by $z\sim 17$ ($t_{\rm Univ}\sim 0.8$ Gyr) 
\citep{Mel23}, the longer time afforded the evolution of the most massive galaxies 
in this model may solve the growing puzzle of how structure formed in the
early Universe.

\begin{acknowledgements}
We are grateful to the anonymous referee and the editor, Christopher Conselice, 
for very helpful comments that have lead to an improved presentation of the results 
in this paper.  MLC's research is supported by the Chinese Academy of Sciences 
President's International Fellowship Initiative grant number 2023VMB0001
and the grant PID2021-129031NB-I00 of the Spanish Ministry of Science (MICIN). 
JJW is supported by the Natural Science Foundation of China (grant No. 12373053), 
the Key Research Program of Frontier Sciences (grant No. ZDBS-LY-7014) of Chinese 
Academy of Sciences, and the Natural Science Foundation of Jiangsu Province 
(grant No. BK20221562).
\end{acknowledgements}

\bibliographystyle{aasjournal} 
\bibliography{agesJWST}

%

\appendix

\section{A. Calculation of the error bars through a maximum likelihood 
algorithm}\label{.maxlh}
The error bars throughout this paper are calculated by identifying the 
portion of parameter space with a $\chi ^2$ value lower than the 
established numbers in $\chi ^2$ statistics (\S \ref{.methods1}). 
Here we describe an alternative way of deriving them based on a 
Bayesian maximum likelihood approach.

We calculate the value of $\chi^2$ for each set of parameters ($\Pi 
\equiv {\Pi _j;\ j=1,...,n_{\rm par}}$) in comparison with the data, 
and we choose the solution with maximum likelihood $\cal L$ with 
respect to a parameter $\Pi _i$ to obtain a value between $x$ and $x+dx$,
such that 
\begin{equation}
{\cal L}(x< \Pi_i <x+dx)=\frac{{\rm Max}[P[{\rm data}|\Pi _i=x;\,(\Pi _j;\ j\ne i)]]dx}
{\int  d\Pi_i \,{\rm Max}[P[{\rm data}|\Pi _i;\,(\Pi _j;\ j\ne i)]]}\,,
\end{equation} 
where we take
\begin{equation}
P({\rm data}|\Pi )=\exp \left(-\frac{\chi _n^2({\rm data}|\Pi )}{2}\right).
\end{equation}
The quantity $\chi _n^2({\rm data}|\Pi )$ is the $\chi ^2$ of the data in 
comparison with the model with the set of parameters $\Pi$ normalizing the 
error bars to obtain $\chi ^2_{\rm red,\, minimum}=1$ (that is, $\chi _n^2
({\rm data}|\Pi )=\frac{N_{dof}\chi ^2(\Pi )}{\chi ^2_{\rm minimum}}$). This 
means assuming that the best fit is a reasonable fit with $\chi ^2_{\rm red,
\, minimum}=1$, and we correct the under/over-estimation of error bars by
multiplying them by a factor $\sqrt{\chi ^2_{\rm red,\, minimum}}$, the same 
assumption that was used in \S \ref{.methods1}. We note that, with this 
normalization of $\chi _n$, we are being more conservative, allowing a wider 
range of values for the parameters when $\chi ^2_{\rm red,\, minimum}>1$, 
which happens in most of our cases.

For our particular situation, among the seven (or six if we fix the redshift) 
parameters, the most interesting one to explore is the age$_{\rm old}$. We 
carry out the calculation for the data of fluxes of the stacked SED of 13 
galaxies (corresponding to Fig. \ref{Fig:sedrstacked}). The probabilities 
for the 10 discrete values of ages we are using are given in 
Table~\ref{Tab:maxlh}. Interpolating from these numbers we find that: 
${\rm age}_{{\rm old,\, stacked}}\ge 3.9\ {\rm Gyr} \ \  (68\% CL)$, 
and $\ge 2.1\ {\rm Gyr} \ \  (95\% CL)$.

\begin{table}
\caption{Probabilities of the discrete values of ages of the old population 
providing a fit of the stacked SED at rest of the 13 galaxies (corresponding 
to Fig. \ref{Fig:sedrstacked}) using two SSPs. The $P_{\rm mlh}$ values are 
calculated through maximum likelihood. $P_{\rm Avni}$ are calculated using 
Equation~(\ref{Avni}).}
\begin{center}
\begin{tabular}{ccccc}
$t$ (Gyr) & $\chi ^2_{\rm red}$ & $P_{\rm mlh}(age_{\rm old}\le t)$  
& $\frac{P_{\chi _n^2, N_{dof}}(age_{\rm old}\le t)}
{P_{\chi _n^2, N_{dof}}(age_{\rm old}\le 11\ {\rm Gyr})}$ &  
$P_{\rm Avni}(age_{\rm old}\le t)$ \\ \hline
0.005 & -- & 0    & 0 & 0 \\
0.025 & 31.25 & $1.1\times 10^{-6}$ & $3.2\times 10^{-5}$ & $2.7\times 10^{-4}$ \\
0.10 & 33.25 & $1.1\times 10^{-6}$ & $3.2\times 10^{-5}$ & $2.7\times 10^{-4}$ \\
0.29 & 32.39 & $1.1\times 10^{-6}$ & $3.2\times 10^{-5}$ & $2.7\times 10^{-4}$ \\ 
0.64 & 27.19 & $1.0\times 10^{-5}$ & $2.2\times 10^{-4}$ & $1.8\times 10^{-3}$ \\
0.90 & 19.57 & $6.7\times 10^{-4}$ & $6.5\times 10^{-3}$ & 0.043 \\
1.4  & 14.58 & 0.011 & 0.058 & 0.247 \\
2.5  & 11.25 & 0.067 & 0.205 & 0.611 \\
5.0  & 7.56 & 0.509 & 0.730 & 0.986 \\
11.0 & 6.34 & 1 & 1 & 1 \\ \hline
\end{tabular}
\end{center}
\label{Tab:maxlh}
\end{table}

\section{B. Weighted average in asymmetrical Gaussian distributions}
\label{.waver}

Here, we carry out the weighted average of $\log_{10}({\rm age}_{\rm old})$.
Assuming that the positive and negative errors are Gaussian (an asymmetrical 
Gaussian when the positive and negative error are different), the probability 
$P$ of each galaxy to have age $x$ is
\begin{equation}
\label{eq:assygauss}
\mathrm{P_i}(x)=\frac{\sqrt{2/\pi}}{\sigma_l+\sigma_r}\times \left\{\begin{array}{l}
\exp \left[-\frac{(x-\tau _i)^2}{2 \sigma_l^2}\right] \quad ; {\rm if}\ x \leqslant \tau _i\\
\exp \left[-\frac{(x-\tau _i)^2}{2 \sigma_r^2}\right] \quad ; {\rm if}\ x>\tau _i\;,
\end{array}\right.
\end{equation}
where the $\tau _i$ means the $i^{\rm th}$ $\mathrm{age_{old}}$, and $\sigma_l$, 
$\sigma _r$ are, respectively, the negative and positive $\mathrm{1 \sigma}$ 
errors. By taking the cumulative product over all distributions, we can obtain 
the distribution of the average age:
\begin{equation}
    \label{eq:prod}
    \mathrm{P_{aver}}(x)=K\times \prod _{i=1}^{n} \mathrm{P_{i}}(x) ,
\end{equation}
where $n$ is the number of galaxies in each redshift bin, and $K$ is a normalization 
constant. Afterwards, by identifying the 0.159, 0.5, and 0.841 quantiles of the 
overall distribution, we obtain the lower limit, median, and upper limit of its 
1-$\sigma$ confidence interval; or 0.023, 0.5, and 0.977 quantiles for 2-$\sigma$. 

In order to take into account the possibility that the error bars do not overlap, 
we add a factor of correction $\sqrt{\frac{\chi ^2}{N-1}}$ for the error bars.

\section{C. SED fit with {\it le PHARE} package}
\label{.lephare}

{\it Le PHARE} is a set of Fortran commands to compute photometric redshifts and to perform SED 
fitting of galaxies, QSOs or stars 
\citep{Arn99,Ilb06}.\footnote{http://www.cfht.hawaii.edu/$\sim $arnouts/lephare.html}
We will use it in this section to show that the results obtained with our own procedure
(\S \ref{.methods1}) are compatible with those obtained with this publically available
package ({\it le PHARE}).

We take the stacked SED at rest of the 13 galaxies (corresponding to Fig. \ref{Fig:sedrstacked}), 
assuming that the photometric redshifts we obtained are approximately correct within their error 
bars. This is indeed already confirmed by comparing our results with those obtained independently 
by \citet{Lab23}. And we have also seen that the stacked SED with only the four galaxies with 
spectroscopic redshift gives approximately the same V-shaped SED. We run the {\it le PHARE} to 
fit the best galaxy template with extinction that fits our data.

{\it Le PHARE} does not use the combination of two SSPs as we have used in \S \ref{.methods1}. Instead,
it adopts models of exponential star formation history of the type 
\begin{equation}
F_{\rm theor.}(\lambda _i)=\frac{L_0}{4\pi d_L(z)^2(1+z)}\times
\end{equation}\[
\int _0^{\rm age}dt\,\left\langle \exp{\left(-\frac{{\rm age}-t}{\tau}\right)}L_{\rm SSP}
(t,[M/H], A_V;\lambda /(1+z))\right\rangle _T
.\]
The stacked SED already sets $z=0$. There are five free parameters here: the amplitude $L_0$, 
metallicity [M/H], age, $\tau $ and $A_V$. The templates for different values of age, $\tau $ 
and [M/H] are taken from  GISELL library with Salpeter IMF \citep{Arn99,Arn02}. The extinction 
is taken from Calzetti's law \citep{Cal00} (\S \ref{.methods1}).
 
\begin{figure}[h]
\vspace{0cm}
\centering
\includegraphics[width=10cm]{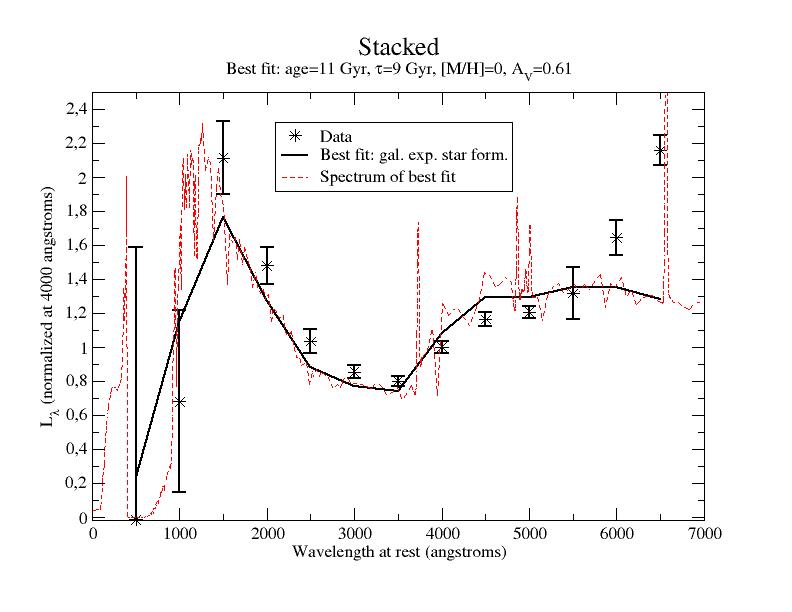}
\vspace{.2cm}
\caption{Best fit using le PHARE package with exponential star formation 
history ($\chi ^2_{\rm red}=14.99$).}
\label{Fig:lephare1}
\end{figure}

The result of the best fit is shown in Fig. \ref{Fig:lephare1}, corresponding to an
age $=11.2^{+1.4}_{-1.5}$ Gyr (68\% CL), $\tau =9$ Gyr, [M/H]=0, $A_V=0.61$ with a 
$\chi ^2=119.9$ (hence, $\chi ^2_{\rm red}=14.99$). This is a worse fit than the one 
we obtained with two SSPs ($\chi ^2=44.4$; $\chi ^2_{\rm red}=6.34$), which indicates 
that our model with two SSPs works much better than an exponential star formation
history in V-shaped SEDs, something that had already been demonstrated by \citet{Lop17}.
Nonetheless, it is remarkable that our best fit corresponds to the same age as that
with two SSPs.

The best fit with an age $<0.5$ Gyr is obtained for age $=64$ Myr, $\tau =0.1$ Gyr,
[M/H]=0, $A_V=2.43$ with a $\chi ^2=174.7$ (hence, $\chi ^2_{\rm red}=21.82$). 
With such a very low $\tau $, it is almost equivalent to a constant star formation 
ratio with an average age $\sim 30$ Myr, which corresponds to our fit in
Fig. \ref{Fig:sedrstacked3}/Middle panel with one SSP of 25 Myr, a very high extinction 
of $A_V=2.48$ and $\chi ^2=170.1$. This is a much worse fit because such a high extinction
erases the Lyman-$\alpha $ peak. 

{\it Le PHARE} also allows for the possibility of fitting a QSO among the different 
templates available with a different continuum power index $\alpha $ and equivalent 
width ($EW$ measured for lines  Ly$\alpha $+NV), without extinction. The best fit 
corresponds to $\chi ^2=155.3$ for $\alpha =-1.25$, $EW=$84 \AA . However, the template 
of the best fit corresponding to synthetic spectra provides fluxes only for 
$\lambda \ge 600$ \AA \ , setting zero flux for lower wavelengths, thus giving a 
false agreement with the point of the bin at 500 \AA \ and considerably reducing 
the $\chi ^2$ with respect to its value when an extrapolation of the template to 
shorter wavelengths is introduced. Moreover, the  extrapolation of Calzetti's law
to wavelengths below 1200 \AA \ is not appropriate. This is critical, because the 
non-detection of the galaxies in the HST filter F435W and the huge dropoff at 
wavelengths shorter than Lyman-$\alpha $ is a clear indication of the incompatibility 
with any QSO SED. Furthermore, it fails to reproduce the area around the Balmer break, 
since the QSO yields blue colors at all wavelengths. A combination of QSO+galaxy might 
improve the fit, but {\it le PHARE} does not allow for the possibility of such fits. 
We have carried out this kind of combination of QSO with 2-SSPs and we have seen that 
adding a QSO component does not improve the fit (see \S \ref{.AGN}).

\section{D. SED fits with the age of galaxies restricted to be younger than that of the Universe}
\label{.ageconstraint}

In Table \ref{Tab:bestfitsc}, we show the results of the best fit with 2SSP, following the method
described in \S \ref{.methods1}, but with the extra constraint that 
${\rm age}_{\rm old}<{\rm age}_{\rm Univ.}(z)$, where ${\rm age}_{\rm Univ.}(z)$ is the age of 
the Universe at redshift $z$ predicted by the standard model. As we can see, the values of 
$\chi ^2_{\rm red}$ are much higher than those shown in Table \ref{Tab:bestfits}, based on
calculations without this constraint: on average, $\chi ^2_{\rm red}$ is 2.2 times larger in 
Table \ref{Tab:bestfitsc} than in Table \ref{Tab:bestfits}. The much worse fits when the 
galaxies' age is constrained to be less than that of the Universe can also be seen
in the optimization of the stacked SED: Left panel of Fig.~\ref{Fig:sedrstacked3}
compared with Fig.~\ref{Fig:sedrstacked}.

The fact that \citet{Lab23} (see also \citet{Bar24} for similar SED fittings of the
JWST galaxies) are able to fit these 13 galaxies with ages younger than that of
the Universe is due to their assumption of a lower, or no, extinction for the young 
population, which is---in our view---inappropriate because the dust extinction 
(if any) should be applied to the whole stellar population, especially the youngest 
component that produces the Far-UV break, given that the dust is more abundant in 
young populations \citep{Mal22}. At a minimum, extinction in the young population 
cannot be lower than the extinction in the old population.

\begin{table*}
\caption{Best fit results with two SSPs and the constraint that
${\rm age}_{\rm old}<{\rm age}_{\rm Univ.}(z)$. Errors represent the limits 
for which the templates of GALAXEV fit the data within 95\% CL ($\equiv 2\sigma $) 
(within the resolution of the templates). The ages are expressed in Gyr. Redshifts 
of id. 13050, 28984, 35900, 39575 are fixed to their spectroscopic [S] values 
(https://dawn-cph.github.io/dja/spectroscopy/nirspec/). References for the
spectroscopic redshifts: \cite{Koc23}: [Koc23]; \cite{Fuj23}: [Fuj23].}
\begin{center}
\begin{tabular}{cccccccccc}
Galaxy ID & $z$ & $\log _{10}[{\rm age_{\rm old}}]$ & $\log _{10}[{\rm age_{\rm young}}]$ & $A_2$ & $[M/H]$ & $A_V$ & $\chi ^2_{\rm red}$ 
\\ \hline
2859 &  $10.48^{+1.32}_{-5.48}$ & $-0.54^{+0.49}_{-1.06}$ & $-2.30^{+2.11}_{-0}$ &  $0.04^{+0.68}_{-0.04}$ &
$+0.4^{+0}_{-0.8}$ & $0.67^{+2.33}_{-0.67}$ & 9.20 \\
7274 &  $10.47^{+1.33}_{-5.47}$ &   $-0.54^{+0.49}_{-1.06}$ & $-2.30^{+2.11}_{-0}$ &  $0.05^{+0.55}_{-0.05}$ &
$+0.4^{+0}_{-0.8}$ & $0^{+2}_{-0}$ & 91.62 \\
11184 &  $7.18^{+0.47}_{-0.39}$ &   $-0.54^{+0.34}_{-0.46}$ & $-1.00^{+0.29}_{-1.30}$ &  $0.48^{+0.16}_{-0.47}$ &
$0^{+0}_{-0.4}$ & $0^{+0.75}_{-0}$ & 11.62 \\
13050 &  $5.62$ [Koc23] & $-1.60^{+1.56}_{-0.22}$ & $-2.30^{+2.11}_{-0.00}$ &  $0^{+0.55}_{-0}$ &
$+0.4^{+0}_{-0.8}$ & $4.26^{+0.84}_{-3.26}$ & 8.31 \\
14924 &  $9.83^{+0.53}_{-0.43}$ &   $-0.54^{+0.21}_{-0.17}$ & $-2.30^{+1.30}_{-0}$ &  $0.04^{+0.26}_{-0.04}$ &
$0^{+0}_{-0.4}$ & $0.23^{+1.27}_{-0.23}$ & 5.71 \\
16624 & $10.20^{+0.88}_{-0.85}$ &  $-0.54^{+0.21}_{-0.46}$ & $-2.30^{+1.30}_{-0}$ &  $0.08^{+0.49}_{-0.07}$ &
$0^{+0}_{-0.4}$ & $0.11^{+0.89}_{-0.11}$ & 9.06 \\ 
21834 &  $10.17^{+1.03}_{-0.53}$ &   $-0.54^{+0.21}_{-0.46}$ & $-2.30^{+1.30}_{-0}$ &  $0.04^{+0.44}_{-0.04}$ &
$+0.4^{+0}_{-0.8}$ & $0.04^{+1.96}_{-0.04}$ & 1.46 \\
25666 &  $7.59^{+2.41}_{-2.59}$ &   $-0.19^{+0.15}_{-1.41}$ & $-2.30^{+2.11}_{-0}$ &  $0.20^{+0.40}_{-0.20}$ &
$+0.4^{+0}_{-0.8}$ & $0.53^{+1.47}_{-0.53}$ & 36.15 \\
28984 &  $7.09$ [S] &   $-0.19^{+0.08}_{-1.41}$ & $-2.30^{+1.76}_{-0}$ &  $0.33^{+0.32}_{-0.33}$ &
$+0.4^{+0}_{-0.8}$ & $1.24^{+0.86}_{-1.24}$ & 17.44  \\
35300 &  $7.77$ [Fuj23]  &   $-0.19^{+0.08}_{-1.41}$ & $-2.30^{+1.76}_{-0.00}$ &  $0.36^{+0.24}_{-0.35}$ &
$+0.4^{+0}_{-0.8}$ & $2.02^{+0.98}_{-2.02}$ & 4.70 \\
37888 &  $7.24^{+2.76}_{-2.24}$ &   $-0.19^{+0.15}_{-1.41}$ & $-2.30^{+2.11}_{-0}$ &  $0.26^{+0.34}_{-0.26}$ &
$+0.4^{+0}_{-0.8}$ & $0.41^{+1.59}_{-0.41}$ & 9.43  \\
38094 &  $8.10^{+0.62}_{-2.10}$ &  $-0.54^{+1.58}_{-0.46}$ & $-2.30^{+2.11}_{-0}$ &  $0.02^{+0.58}_{-0.02}$ &
$0^{+0}_{-0.4}$ & $0.42^{+1.58}_{-0.42}$ & 142.92  \\
39575 &  $7.99$ [Fuj23] &  $-0.54^{+0.21}_{-1.06}$ & $-2.30^{+1.30}_{-0}$ &  $0.56^{+0.28}_{-0.55}$ &
$-0.4\pm 0$ & $1.44^{+0.66}_{-1.44}$ & 4.21  \\
\end{tabular}
\end{center}
\label{Tab:bestfitsc}
\end{table*}

\end{document}